\documentclass[seceq]{ptptex}
\usepackage{wrapft}

\newcommand{\pd}[3]{\left( \frac{\partial #1}{\partial #2} \right)_{#3}}

\newcommand{\integ}[1]{\int d#1 \hspace*{1mm}}

\usepackage{amssymb}
\usepackage[dvips]{graphicx}
\usepackage{booktabs}
\usepackage{tabularx}
\usepackage{graphicx,color} 
\usepackage{wrapft}

\pubinfo{}
\notypesetlogo
\title{Nonequilibrium Peierls Transition}

\author{
Shigeru \textsc{Ajisaka}\footnote{Email: g00k0056@suou.waseda.jp}, Hisashi \textsc{Nishimura}

Shuichi \textsc{Tasaki}
and Ichiro \textsc{Terasaki}
}

\inst{
Advanced Institute for Complex Systems
and 
Department of Applied Physics,\\
School
of Science and Engineerings,
Waseda University,\\
Tokyo 169-8555,
Japan
}

\abst{
The nonequilibrium phase transition of an open Takayama-Lin Liu-Maki chain coupled with two reservoirs
is investigated by combining a mean-field approximation and a formula characterizing nonequilibrium steady states,
which is obtained from the algebraic field theoretical 
approach to nonequilibrium statistical mechanics.
When the bias voltage is chosen as a control parameter, the phase transition between ordered and normal phases is 
found to be 
of 
first or second order. The
current-voltage 
characteristics are S-shaped in some parameter region. 
In contrast, when the current is chosen as a control parameter, all the 
nontrivial solutions of the self-consistent equation are found to be stable. In this case, the phase transition 
between the ordered and normal phases is always
of 
second order, and negative differential conductivity appears at 
low temperature.
}

\begin{document}
\maketitle

\section{Introduction}
One of the main interests in mesoscopic physics is a full understanding of the nonequilibrium properties
of small quantum systems coupled with large reservoirs. When the reservoirs have different temperatures
and/or chemical potentials, the whole system can be in a steady state with constant particle and 
energy flows, i.e. a nonequilibrium steady state (NESS). One of promising approaches in 
dealing with such systems is based on algebraic quantum field theory\cite{BR,Haag,LecMath1880}. 
For example, with the aid of the algebraic approach, Pusz and Woronowicz rigorously derived 
Carnot's formula\cite{PW,BR} and Ojima {\it et al.} proved the positivity of the (relative) entropy production 
rate\cite{Ojima}, both for systems of infinite degrees of freedom.
In addition, for the XY model, a NESS was rigorously constructed\cite{HoAraki}.
Recently, starting from Ruelle's work on scattering-theoretical characterizations of NESS\cite{Rue} and 
Jak\v{s}i\'c-Pillet's investigation into entropy production\cite{JP}, the algebraic approach to NESS has 
been extensively developed [see 
Refs.\citen{JPRev,FrolichEQ,TasakiRev,TasakiRev2,TasakiTherm} 
and references therein]. Currently, 
linear response theories\cite{LinRes}, 
thermodynamic
properties\cite{TasakiTherm,Therm} and the Landauer-B\"uttiker 
formula\cite{AJPP,TakTasFree} are investigated, 
in addition to various other 
aspects\cite{Others}.

Hereafter, we consider a quantum system coupled with two free fermionic reservoirs $L$ and $R$ described 
by annihilation operators $a_{{\bf k}\sigma}$ and $b_{{\bf k}\sigma}$, respectively, where ${\bf k}$ refers 
to the wave number and $\sigma$ to the spin. 
In this case, if the reservoirs are 
initially
set to be in different equilibria, the whole system is shown to 
approach a NESS in the long time limit provided that the {\it incoming} fields 
$\alpha_{{\bf k}\sigma}$ of $a_{{\bf k}\sigma}$ and $\beta_{{\bf k}\sigma}$ of $b_{{\bf k}\sigma}$
are complete\cite{TakTasFree,TasakiRev2}. 
The NESS so obtained
can be characterized as a state satisfying Wick's theorem with respect to $\alpha_{{\bf k}\sigma}$ 
and $\beta_{{\bf k}\sigma}$ and having the two-point functions:
\begin{equation}\label{ChNESS}
\langle \alpha_{{\bf k}\sigma}^\dag \alpha_{{\bf k}'\sigma'}
\rangle_\infty= f_L(\hbar\omega_{k L}) \delta_{\sigma\sigma'}\delta({\bf k}-{\bf k}') \ ,
\ \
\langle \beta_{{\bf k}\sigma}^\dag \beta_{{\bf k}'\sigma'}
\rangle_\infty= f_R(\hbar\omega_{k R}) \delta_{\sigma\sigma'}\delta({\bf k}-{\bf k}') \ ,
\end{equation}
where 
$\langle \cdots\rangle_\infty$ stands for the average with respect to the
NESS,
$\hbar\omega_{k \nu}$ is the single-particle energy of wave number $\bf k$,  
$f_\nu(x)\equiv 1/(e^{(x-\mu_\nu)/T_\nu}+1)$ is the Fermi distribution function, $T_\nu$ 
is the initial temperature and $\mu_\nu$ is the initial chemical potential of the reservoir $\nu=L,R$.
Formally, this can be understood as follows:\footnote{The very proof of the existence of the limits
requires rigorous and careful arguments\cite{TasakiRev2}.}
Let $\rho_0$ be an initial density matrix, where the two reservoirs are in distinct equilibria, and
$\rho_{\infty}$ be that of the NESS, 
then, $\lim_{t\to +\infty}e^{-iHt/\hbar}\rho_0e^{iHt/\hbar}=\rho_\infty$ and,
e.g., ${\rm Tr}\{a_{{\bf k}\sigma}^\dag a_{{\bf k}'\sigma'}\rho_0\}
=f_L(\hbar\omega_{k L}) \delta_{\sigma\sigma'}\delta({\bf k}-{\bf k}')$. 
As the incoming field $\alpha_{{\bf k}\sigma}$ is given by $\lim_{t\to +\infty}
e^{i\omega_{kL}(-t)}e^{iH(-t)/\hbar}a_{{\bf k}\sigma}e^{-iH(-t)/\hbar}=\alpha_{{\bf k}\sigma}$,
one obtains the desired relation:
\begin{eqnarray}
&&f_L(\hbar\omega_{k L}) \delta_{\sigma\sigma'}\delta({\bf k}-{\bf k}')
={\rm Tr}\{a_{{\bf k}\sigma}^\dag a_{{\bf k}'\sigma'}\rho_0\}e^{i(\omega_{kL}-\omega_{k'L})t}
\nonumber\\
&&
={\rm Tr}\Big\{
\{e^{i{H\over \hbar}(-t)}a_{{\bf k}\sigma}e^{-i{H\over \hbar}(-t)}e^{i\omega_{kL}(-t)}\}^\dag 
e^{i{H\over \hbar}(-t)}a_{{\bf k}'\sigma'}e^{-i{H\over \hbar}(-t)}e^{i\omega_{k'L}(-t)}
e^{-i{H\over \hbar}t}\rho_0 e^{i{H\over \hbar}t}
\Big\}
\nonumber\\
&&\to 
{\rm Tr}\{\alpha_{{\bf k}\sigma}^\dag \alpha_{{\bf k}'\sigma'}\rho_\infty\}
\equiv \langle \alpha_{{\bf k}\sigma}^\dag \alpha_{{\bf k}'\sigma'}\rangle_\infty
\quad ({\rm as \ } t\to +\infty) \ .
\end{eqnarray}
As noted by Blanter and B\"uttiker\cite{BButtiker} [cf. Eqs.(29) and (36) in their paper], the NESS 
characterization (\ref{ChNESS}) can be a starting point for the Landauer-B\"uttiker approach to transport
properties of
mesoscopic circuits. Note also that (\ref{ChNESS}) can be applied even to systems 
with interacting fermions 
if 
the incoming fields are complete\cite{TasakiRev2}. 
Indeed, Katsura successfully derived a NESS from (\ref{ChNESS}) for a solvable model of the Kondo effect\cite{Katsura}.

Since eq.(\ref{ChNESS}) fully characterizes NESSes of noninteracting fermions,
it is natural to consider a mean-field approximation based on (\ref{ChNESS})
for a NESS of interacting fermions. 
Based on this view, we investigated a
NESS of an Aharonov-Bohm ring with a quantum dot within a mean-field approximation and obtained a differential 
conductivity consistent with numerical renormalization group analysis and experiments\cite{TakTasABDot}. 
Here, with the aid of a similar nonequilibrium mean-field approximation, 
we study the nonequilibrium phase transition in the Takayama-Lin Liu-Maki chain\cite{TLM} (TLM chain) embedded between 
two infinitely extended reservoirs. 
The TLM chain is a continuum limit of a lattice model (the SSH lattice) 
for polyacetylene proposed by Su, Schrieffer and Heeger\cite{SSH} and describes the charge density wave commensurate 
with the lattice.

We employ the TLM chain as a representative example of systems with phase transitions. However, since the mean 
field approximations of the TLM chain, 
a superconductors, a 
1D extended Hubbard lattice of spinless 
fermions and the Jordan-Wigner-transformed XXZ model are equivalent, the present analysis would provide some 
insight into nonequilibrium properties of 
various
interacting systems, such as the current-induced suppression of the charge 
order observed in some $\theta$-type BEDT-TTF organic conductors\cite{Inagaki,Sawano1,Wata,Sawano2} and the negative 
differential conductivity recently reported for the XXZ model\cite{Casati} and some strongly correlated 
systems\cite{Egger,Oka,Casati2}.

The rest of this paper is arranged as follows.
In Sec.~2, we introduce a finite TLM chain coupled with two infinitely extended reservoirs. In Sec.~3, a mean-field 
approximation based on (\ref{ChNESS}) is discussed. The averaged lattice distortion serves as an order parameter, and 
its self-consistent equation is obtained by averaging the equation of motion of the lattice distortion with respect 
to a nonequilibrium steady state. 
In Sec.~4, the self-consistent equation, current and stability conditions are explicitly derived in case where the TLM 
chain is long enough and the order parameter is spatially uniform. 
In Sec.~5, possible phases are discussed in detail when the chain-reservoir couplings are symmetric. 
When the bias voltage is chosen as 
one of the 
control parameters, the phase transition between ordered and normal 
phases could be 
of
first or second order depending on 
the bias voltage and temperature. 
At low temperature, the current-voltage
characteristics are S-shaped.
For some bias-voltages, the temperature dependence of the order parameter is found to be similar to that for 
the nonequilibrium superconducting phase induced by excess quasiparticles\cite{Scalapino1,Scalapino2}.
In contrast, when the current is chosen as 
one of the
control parameters, 
the self-consistent equation has a unique stable solution
and the phase transition between the ordered and normal phases is always 
of
second order. Negative differential conductivity appears when the temperature is lower than a certain threshold value.
In Sec.~6, after a summary of the paper is given, the self-consistent equation for the open TLM chain is compared with
that for the nonequilibrium superconductor obtained by Scalapino {\it et al.}\cite{Scalapino1}. 
Then, on the basis of the similarity of mean-field approximations of the open TLM chain 
and an open 1D extended Hubbard lattice, the experimental results for some $\theta$-type BEDT-TTF organic 
conductors\cite{Inagaki,Sawano1,Wata,Sawano2} as well as the negative differential conductivity found in an open XXZ 
model\cite{Casati} are qualitatively discussed within the scope of the present analysis.
In Appendix A, an open TLM chain is derived from an open SSH lattice. 
In Appendix B, normal-mode operators are explicitly given. In Appendix C, 
we discuss
the relationship between the average chemical
potential and the Coulomb energy. 
In Appendix D, Green functions necessary for deriving the self-consistent equation are provided.
In Appendix E, the stability of the nontrivial phases at zero temperature is discussed.
In Appendix F, the Ginzburg-Landau expansion for the self-consistent equation is given.

\section{Open TLM Model}

The system in question consists of a finite TLM chain and two free electron reservoirs. In terms of 
the quantized local lattice distortion $\Delta(x)$ and the two-component electron field $\Psi_\sigma(x)$
\begin{equation}
\Psi_\sigma(x)\equiv \left(
\begin{matrix}
d_\sigma(x)\cr e_\sigma(x)
\end{matrix}
\right)
\ ,
\end{equation}
the Hamiltonian of the TLM chain is given by\cite{TLM}
\begin{eqnarray}\label{HamHS}
H_S &=&
\sum_{\sigma} \int_0^\ell dx
\Psi_\sigma^\dag(x)
\left[-i \hbar v
\sigma_y\frac{\partial }{\partial x}+ 
\Delta (x) \sigma_x
\right]
\Psi_\sigma(x)
\nonumber \\
&&+
{1\over 2\pi \hbar v \lambda}\int_0^\ell dx 
\left[\Delta(x)^2
+{1\over \omega_0^2}\Pi(x)^2
\right]
\ ,
\end{eqnarray}
where
$\ell$ is the length of the system, $v$ is the Fermi velocity, $\sigma_x$ and 
$\sigma_y$ are the $x$ and $y$ components of Pauli matrices, $\lambda$ is the 
dimensionless coupling constant, $\omega_0$ is the 
phonon frequency and $\Pi(x)$ corresponds to the momentum conjugate to $\Delta(x)$. 
Nonvanishing equal-time commutation relations among those operators are
\begin{eqnarray}
&&\{d_\sigma(x), d_{\sigma'}(y)^\dag\}=
\{e_\sigma(x), e_{\sigma'}(y)^\dag\}=\delta_{\sigma,\sigma'}\delta(x-y) \ ,
\\
&&\left[\Delta(x),\Pi(y)\right]=i \hbar^2 \pi \lambda v \omega_0^2 \delta(x-y) \ ,
\end{eqnarray}
where $\{A, B\}=AB+BA$ and $[A,B]=AB-BA$. As the system is finite, electron waves are reflected back at the 
edges, and the following boundary condition is imposed:
\begin{equation}\label{TLMboundary}
d_\sigma(0)=0 \ , \ \ \ e_\sigma(\ell)=0 \ .
\end{equation}
The fields $d_\sigma$ and $e_\sigma$ correspond to electrons
at even and odd sites, respectively, of a finite SSH lattice (cf. Appendix~A).
Note that, instead of $d_\sigma$ and $e_\sigma$, the original work\cite{TLM} uses the right- 
and left-moving electron fields 
$\psi_{R \sigma}(x)$ and $\psi_{L \sigma}(x)$, respectively,
\begin{equation}
\psi_{R \sigma}(x)={1\over \sqrt{2}}\{d_\sigma(x)-ie_\sigma(x) \}
\ , \quad
\psi_{L \sigma}(x)={1\over \sqrt{2}}\{e_\sigma(x)-id_\sigma(x) \}
\ .
\end{equation}

The reservoirs are described by
\begin{equation}\label{HamB}
H_B=\sum_{\sigma}\int d{\bf k} \{ \hbar\omega_{kL} a_{{\bf k}\sigma}^\dag a_{{\bf k}\sigma} +
\hbar \omega_{kR} b_{{\bf k}\sigma}^\dag b_{{\bf k}\sigma}\} \ ,
\end{equation}
where $a_{{\bf k}\sigma}$ and $b_{{\bf k}\sigma}$ stand for the annihilation operators of electrons with 
wave number $\bf k$ and spin $\sigma$ in the left and right reservoirs, respectively, and
$\hbar \omega_{k\nu}$ ($\nu=L,R$) 
are their energies 
measured from the zero-bias chemical potentials at absolute 
zero 
temperature. The nonvanishing anticommutation relations
among them are $\{a_{{\bf k}\sigma},a_{{\bf k}'\sigma'}^\dag\}
=\{b_{{\bf k}\sigma},b_{{\bf k}'\sigma'}^\dag\}=\delta_{\sigma\sigma'}\delta({\bf k}-{\bf k}')$.
The chain-reservoir interaction is assumed to be
\begin{equation}\label{HamInt}
V=\sum_{\sigma}\int d{\bf k} \ 
\bigg\{ \hbar v_{\bf k}
e_\sigma^\dag(0) 
a_{{\bf k}\sigma} + \hbar
w_{\bf k}
d_\sigma^\dag(\ell) 
b_{{\bf k}\sigma} + (h.c.) \bigg\} \ ,
\end{equation}
where $v_{\bf k}$ and $w_{\bf k}$ stand for the coupling matrix elements. 

The Hamiltonian of the whole system is given by
\begin{equation}
H=H_S+V+H_B
\ .
\label{TotalH}
\end{equation}
As will be discussed in Appendix~A, an open TLM chain described by $H$
corresponds to an open SSH lattice 
that couples with the reservoirs through the end sites
and the number of whose sites is a multiple of four.

From (\ref{TotalH}), the lattice distortion $\Delta$ is found to obey the following equation of motion: 
\begin{eqnarray}
\frac{\partial \Delta(x,t)}{\partial t}&=&{1\over i\hbar}[\Delta(x,t),H]
=\Pi(x,t)
\nonumber
\\
\frac{\partial^2 \Delta(x,t)}{\partial t^2}&=&
\frac{\partial \Pi(x,t)}{\partial t}
=-\omega_0^2
\left(\Delta(x,t)+\pi \hbar v \lambda
\sum_{\sigma}
\Psi_\sigma^\dag(x,t) \sigma_x \Psi_\sigma(x,t)
\right)\ .
\label{DeltaEq}
\end{eqnarray}
Eqs.(\ref{TotalH}) and (\ref{DeltaEq}) are our starting points. 

\section{NESS Mean-Field Approximation}

In this section, we describe
a procedure for evaluating the NESS averages of the electron variables for 
the TLM chain and 
derive the self-consistent equation for the NESS average ${\overline \Delta}(x)
\equiv \langle \Delta(x)\rangle_\infty$ of the lattice distortion, which serves
as the order parameter of the Peierls transition.

The nonequilibrium steady state under the mean-field approximation
is characterized by (\ref{ChNESS}) where $\alpha_{{\bf k}\sigma}$ 
and $\beta_{{\bf k}\sigma}$ are the incoming fields of $a_{{\bf k}\sigma}$ and  $b_{{\bf k}\sigma}$ with
respect to the mean-field Hamiltonian:
\begin{eqnarray}
H_{\rm \small MF}&=&H_S^{\rm \small MF}+V+H_B
\\
H_S^{\rm \small MF}&\equiv&
\sum_{\sigma} \int_0^\ell dx
\Psi_\sigma^\dag(x)
\left[-i \hbar v
\sigma_y\frac{\partial }{\partial x}+ 
{\overline\Delta} (x) \sigma_x
\right]
\Psi_\sigma(x) \ .
\label{NESSMF1}
\end{eqnarray}
Namely, they are defined as the solution of
\begin{eqnarray}
&&{1\over \hbar} [\alpha_{{\bf k}\sigma}, H_{\rm \small MF} ]=\omega_{kL}\alpha_{{\bf k}\sigma}
\ ,\ \ \ 
e^{iH_{\rm \small MF}t/\hbar} a_{{\bf k}\sigma} e^{-iH_{\rm \small MF}t/\hbar} \ e^{i\omega_{kL} t} 
\to \alpha_{{\bf k}\sigma}
\
(t\to -\infty)
\label{InCom1}
\\
&&{1\over \hbar} [\beta_{{\bf k}\sigma}, H_{\rm MF} ]=\omega_{kR}\beta_{{\bf k}\sigma}
\ ,\ \ \ 
e^{iH_{\rm \small MF}t/\hbar} b_{{\bf k}\sigma} e^{-iH_{\rm \small MF}t/\hbar} \ e^{i\omega_{kR} t} 
\to \beta_{{\bf k}\sigma} \
(t\to -\infty) \
\ .
\label{InCom2}
\end{eqnarray}
Since the mean-field Hamiltonian $H_{\rm\small MF}$ is bilinear with respect to the electron creation/annihilation 
operators, the incoming fields are linear combinations of $a_{{\bf k}\sigma}$, $b_{{\bf k}\sigma}$, and $\Psi_\sigma(x)$.
As shown in Appendix~\ref{ap.normal}, the incoming fields are fully determined by (\ref{InCom1}) and (\ref{InCom2}).
Conversely, the original operators can be represented by the incoming fields. For example, we have
\begin{eqnarray}\label{PsiNormal}
\Psi_\sigma(x)
=\int d{\bf k} 
\Big\{
v_{\bf k}{h(x;\omega_{kL}) \over \Lambda_-(\omega_{kL})^*} \alpha_{{\bf k}\sigma}+
w_{\bf k}{{\widetilde h}(x;\omega_{kR})\over \Lambda_-(\omega_{kR})^*}\beta_{{\bf k}\sigma}
\Big\}
\ ,
\end{eqnarray}
where $\Lambda_-(\omega)$, $h(x;\omega)$ and ${\widetilde h}(x;\omega)$ are auxiliary functions
given by
\begin{eqnarray}
\Lambda_-(\omega)&=&1-\xi_-(\omega)g_{--}(0,0;\omega)-\eta_-(\omega)g_{++}(\ell,\ell;\omega)
\nonumber \\
&+&\xi_-(\omega)\eta_-(\omega)
\{g_{++}(\ell,\ell;\omega)g_{--}(0,0;\omega)-g_{+-}(\ell,0;\omega)g_{-+}(0,\ell;\omega)
\}
\ ,
\\
h(x;\omega)&=& G(x,0;\omega)\binom{0}{1} \{1-g_{++}(\ell,\ell;\omega)\eta_+(\omega)\}
\nonumber\\
&&+G(x,\ell;\omega)\binom{1}{0} g_{+-}(\ell,0;\omega)\eta_+(\omega)
\ ,
\\
{\widetilde h}(x;\omega)&=& 
G(x,0;\omega)\binom{0}{1} g_{-+}(0,\ell;\omega)\xi_+(\omega)
\nonumber\\
&&+G(x,\ell;\omega)\binom{1}{0} \{1-g_{--}(0,0;\omega)\xi_+(\omega)\}
\ ,
\\
\xi_\pm(\omega)&=& \integ{{\bf k}'}\frac{|v_{{\bf k}'}|^2}{\omega-\omega_{k'L}\pm i0}
,\ \ \ 
\eta_\pm(\omega)= \integ{{\bf k}'}\frac{|w_{{\bf k}'}|^2}{\omega-\omega_{k'R}\pm i0}
\ .
\end{eqnarray}
And the Green function
$G$ for the finite TLM chain
and, equivalently, its components $g_{\sigma\sigma'}$ ($\sigma ,\sigma'=\pm$)
are defined as a solution of
\begin{eqnarray}
&&
G(x,y;\omega)\equiv
\left(\begin{matrix}
g_{++}(x,y;\omega) &g_{+-}(x,y;\omega)\cr
g_{-+}(x,y;\omega) &g_{--}(x,y;\omega)\cr
\end{matrix}\right)
\ ,
\label{GreenFt}
\\
&&\left[-i \hbar v
\sigma_y\frac{\partial }{\partial x}+ 
{\overline\Delta}(x) \sigma_x
\right]G(x,y:\omega)
=\hbar \omega G(x,y;\omega)-\hbar {\bf 1}\delta(x-y)
\ ,
\label{GreenFt1}
\\
&&
g_{++}(0,y;\omega)=g_{+-}(0,y;\omega)=g_{-+}(\ell,y;\omega)=g_{--}(\ell,y;\omega)=0
\ .
\label{GreenFt2}
\end{eqnarray}
For the derivation of incoming-field operators,
see Appendix~\ref{ap.normal}.
Then, the mean-field NESS is given as a state satisfying Wick's theorem
with respect to $\alpha_{{\bf k}\sigma}$ and $\beta_{{\bf k}\sigma}$ with the two-point 
functions (\ref{ChNESS}).
Note that, although the Green function (and, thus, $\Lambda_-$, $h$ and $\widetilde h$)
diverges as a function of $\omega$ at eigenvalues of the 
differential operator in the left-hand side of (\ref{GreenFt1}),
the integrand of (\ref{PsiNormal}) remains finite even when $\omega_{kL}$ or $\omega_{kR}$ is 
equal to one of the eigenvalues.
The NESS average of any electron variable consisting of $\Psi_\sigma$ can be calculated from 
(\ref{ChNESS}) and (\ref{PsiNormal}).

The self-consistent equation for the order parameter ${\overline \Delta}(x)$ is derived from the
equation of motion (\ref{DeltaEq}) for the lattice distortion.
Because of the time-independence of ${\overline \Delta}(x)$, (\ref{DeltaEq}) leads to
\begin{eqnarray}\label{Self1}
0={-1\over \omega_0^2}\frac{\partial^2 {\overline\Delta}(x)}{\partial t^2}&=&
{\overline\Delta}(x)+\pi \hbar v \lambda
\sum_{\sigma}\langle
\Psi_\sigma^\dag(x) \sigma_x \Psi_\sigma(x)\rangle_\infty
\ .
\end{eqnarray}
Using (\ref{ChNESS}) and (\ref{PsiNormal}), $\displaystyle
\int d{\bf k}|v_{\bf k}|^2F(\omega_{kL})=\int_{-\infty}^\infty d\omega
F(\omega) \ {\rm Im}\xi_-(\omega)/\pi$ and a similar formula for $\eta_-(\omega)$,
the self-consistent equation (\ref{Self1}) reads as
\begin{eqnarray}\label{Self2}
&&
{2\over\pi}
\int_{-\infty}^\infty d\omega 
\Biggl\{{\rm Im} \xi_-(\omega)
{
h(x;\omega)^\dag \sigma_x
h(x;\omega) 
\over |\Lambda_-(\omega)|^2} f_L(\hbar\omega)
\nonumber\\
&&\mskip 140mu +{\rm Im} \eta_-(\omega)
{
{\widetilde h}(x;\omega)^\dag
\sigma_x {\widetilde h}(x;\omega)
\over |\Lambda_-(\omega)|^2} f_R(\hbar\omega)
\Biggr\} 
=-
{{\overline\Delta}(x)\over \pi \hbar v \lambda}
\ ,
\end{eqnarray}
where we use the convention ${\rm Im}\xi_-(\omega)=0$ (${\rm Im}\eta_-(\omega)=0$) for $\omega$ 
outside the range of $\omega_{kL}$ ($\omega_{kR}$).
Equation (\ref{Self2}) is the self-consistent equation for the order parameter
${\overline\Delta}(x)$.

\section{Spatially Uniform Phase}

\subsection{Self-consistent equation}

Hereafter, we consider the cases where 
the chain is half-filled; namely, the zero-bias chemical potentials are located at 
the band centre of the TLM chain. 
Then, in order to prevent an increase in electrostatic energy, the chemical potentials 
of the reservoirs should be chosen such that $\mu_L=-\mu_R=-eV/2$ where $V$ is the
bias voltage and $e$ is the elementary charge (for details, see Appendix~\ref{ap.chem}).
It is well known\cite{TLM} that the energy cutoff $\hbar\omega_c$
($\hbar\omega_c\gg T, e|V|$) is necessary for the
TLM model, and the integration 
interval of (\ref{Self2}) should be replaced with $(-\omega_c,\omega_c)$.

The local state of the TLM chain within a certain region of the boundary 
is affected by the existence of the reservoirs. 
However, 
if the chain-reservoir interaction 
is not too strong and if the size of this region is much smaller than 
the length of the chain,
boundary effects might be neglected. From this observation, 
we study the uniform phase
where the order parameter is independent of the coordinate:
${\overline\Delta}(x)={\overline\Delta}$.
In this subsection, we derive the self-consistent equation for ${\overline\Delta}$
when the TLM chain is sufficiently long.

When the order parameter is spatially uniform,
the Green function defined by (\ref{GreenFt1}) can be easily obtained
for any real number $\omega$ and we have 
(for its complete expression,
see Appendix~\ref{ap.Green})
\begin{eqnarray}
&&g_{++}(x,\ell;\omega)=g_{--}(\ell-x,0;\omega)=-{\hbar\omega\sin \kappa x\over v D(\omega)}
\ ,
\\
&&g_{-+}(x,\ell;\omega)=g_{+-}(\ell-x,0;\omega)=-{\hbar\kappa v\cos \kappa x+{\overline\Delta}\sin
\kappa x \over v D(\omega)} \ ,
\end{eqnarray}
where $\kappa=\sqrt{(\hbar\omega)^2-{\overline\Delta}^2}/(\hbar v)$ and 
$D(\omega)=\hbar\kappa v\cos \kappa \ell+{\overline\Delta}\sin \kappa \ell$.
Then, the second term of the left-hand side of (\ref{Self2}) reads
as
\begin{eqnarray}\label{Aux}
&&
\int_{|\omega|<\omega_c} d\omega 
{\rm Im} \eta_-(\omega)
{
{\widetilde h}(x;\omega)^\dag
\sigma_x {\widetilde h}(x;\omega)
\over |\Lambda_-(\omega)|^2} f_R(\hbar\omega) 
\nonumber\\
&&=\int_{|\omega|<\omega_c} d\omega 
{
{\rm Im} \eta_-(\omega)\
f_R(\hbar\omega)
\over |v^2D(\omega)\Lambda_-(\omega)|^2}
\Big\{\hbar^2 \kappa v \omega (v^2-|\xi_-(\omega)|^2)\sin 2\kappa x
\nonumber\\
&&~~~~~~~~~~~~~~~~~~-(\hbar\omega v^2{\overline\Delta}+2v(\hbar\omega)^2{\rm Re}\xi_-(\omega)
+\hbar\omega{\overline\Delta}|\xi_-(\omega)|^2)\cos 2\kappa x 
\nonumber\\
&&~~~~~~~~~~~~~~~~~~+\hbar v^2
{\overline\Delta}\Big(\omega+{2{\overline\Delta}\over \hbar v}{\rm Re}\xi_-(\omega)
+{\omega\over v^2}|\xi_-(\omega)|^2\Big)\Big\}
{\rm sgn}(|\hbar\omega|-|{\overline\Delta}|)
\ ,
\end{eqnarray}
where the denominator is a function of $\omega$ and $\kappa\ell$:
$|v^2D(\omega)\Lambda_-(\omega)|^2\equiv |{\widetilde\Lambda}_-(\omega,\kappa\ell)|^2$ with
\begin{eqnarray}
{\widetilde\Lambda}_-(\omega,\theta)&=&
\hbar\kappa v\big(v^2-\xi_-(\omega)\eta_-(\omega)\big)\cos\theta \nonumber\\
&&+\big(v^2{\overline\Delta}+v\hbar\omega \{\xi_-(\omega)+\eta_-(\omega)\}+{\overline\Delta}
\xi_-(\omega)\eta_-(\omega)\big)\sin\theta
\ .
\end{eqnarray}

Now we show that (\ref{Aux}) is simplified when the TLM chain is sufficiently long.
It is easy to see that if $|\hbar\omega|<|{\overline\Delta}|$, $\kappa$ is purely imaginary, 
$|D(\omega)|^2{\widetilde h}(x;\omega)^\dag
\sigma_x {\widetilde h}(x;\omega)\sim e^{2|\kappa|x}$
and $|D(\omega)\Lambda_-(\omega)|^2
\sim e^{2|\kappa|\ell}$. Then, the integrand of (\ref{Aux})
is on the order of
\begin{eqnarray}
&&
{
{\widetilde h}(x;\omega)^\dag
\sigma_x {\widetilde h}(x;\omega)
\over |\Lambda_-(\omega)|^2} \sim e^{-2|\kappa|(\ell-x)} 
\ ,
\end{eqnarray}
which is negligible for large $\ell$ unless $x$ is near 
the chain ends. Hence, the contribution to (\ref{Aux}) from the interval 
$0<\omega<|{\overline \Delta}|/\hbar$ is negligible. 
This implies that $|{\overline\Delta}|$ corresponds to the electronic energy 
gap as in the equilibrium case.

On the other hand, if $|\hbar\omega|>|{\overline\Delta}|$, $\kappa$ is real.
Then,
with the aid of the Fourier expansion of
$1/|{\widetilde\Lambda}_-(\omega,\theta)|^2$,
$$
{1 \over |{\widetilde\Lambda}_-(\omega,\theta)|^2}=\sum_{n=-\infty}^\infty {e^{2ni\theta}\over \zeta_n(\omega)}
\ ,
$$
where $1/\zeta_n(\omega)$ is the Fourier coefficient,
one can show that the first and second terms of the right-hand side of (\ref{Aux}) 
are negligible, and that the denominator in the third term is replaced with $\zeta_0(\omega)$
provided $\ell$ is large and $x$ is far from the chain ends.
For example, the first term of (\ref{Aux}) becomes
\begin{eqnarray}\label{AuxAux}
&&\int_{|{\overline\Delta}|/\hbar<|\omega|<\omega_c} d\omega 
{
{\rm Im} \eta_-(\omega)\
f_R(\hbar\omega)
\over |v^2D(\omega)\Lambda_-(\omega)|^2}
\hbar^2 \kappa v \omega (v^2-|\xi_-(\omega)|^2)\sin 2\kappa x
\nonumber\\
&&
=\sum_{n=-\infty}^\infty 
\int_{|{\overline\Delta}|/\hbar<|\omega|<\omega_c} {d\omega\over 2i} 
{H(\omega)\over \zeta_n(\omega)} \{e^{2i\kappa(n\ell+x)}
-e^{2i\kappa(n\ell-x)}\}
\ ,
\end{eqnarray}
where
$H(\omega)={\rm Im} \eta_-(\omega)f_R(\hbar\omega)\hbar^2 \kappa v \omega (v^2-|\xi_-(\omega)|^2)$.
When $x$ is far from the chain ends, $n\ell\pm x={\rm O}(\ell)$ and (\ref{AuxAux}) 
is negligible for large $\ell$ thanks to the Riemann-Lebesgue lemma. In short, (\ref{Aux}) is found to be
\begin{eqnarray}\label{Aux2}
&&
\int_{|\omega|<\omega_c} d\omega 
{\rm Im} \eta_-(\omega)
{
{\widetilde h}(x;\omega)^\dag
\sigma_x {\widetilde h}(x;\omega)
\over |\Lambda_-(\omega)|^2} f_R(\hbar\omega) 
\nonumber\\
&&=\hbar v^2{\overline\Delta}\int\limits_{{|{\overline\Delta}|\over\hbar}<|\omega|<\omega_c} d\omega 
{{\rm Im} \eta_-(\omega)\over \zeta_0(\omega)}
f_R(\hbar\omega)
\Big(\omega+{2{\overline\Delta}\over \hbar v}{\rm Re}\xi_-(\omega)
+{\omega\over v^2}|\xi_-(\omega)|^2\Big)
\ .
\end{eqnarray}

By a similar argument, when the TLM chain is sufficiently long, 
the self-consistent equation (\ref{Self2}) 
gives the desired equation for $\overline{\Delta}$
\begin{eqnarray}\label{SelfUni1a}
&& {\overline\Delta}=0 \quad {\rm or}
\\
&&
{-1\over \lambda}=
S({\overline\Delta},V,T_L,T_R)
\ ,
\label{SelfUni1b}
\end{eqnarray}
where 
\begin{eqnarray}\label{SelfUni2}
&&
{S({\overline\Delta},V,T_L,T_R)\over 2 \hbar^2 v^3} \equiv
\int\limits_{{|{\overline\Delta}|\over\hbar}<|\omega|<\omega_c} d\omega 
\Biggl\{
{{\rm Im} \xi_-(\omega)\over \zeta_0(\omega)}
\Big(\omega+{2{\overline\Delta}\over \hbar v}{\rm Re}\eta_-(\omega)
+{\omega\over v^2}|\eta_-(\omega)|^2\Big)
f_L(\hbar\omega)
\nonumber\\
&&\mskip 140mu 
+{{\rm Im} \eta_-(\omega)\over \zeta_0(\omega)}
\Big(\omega+{2{\overline\Delta}\over \hbar v}{\rm Re}\xi_-(\omega)
+{\omega\over v^2}|\xi_-(\omega)|^2\Big)
f_R(\hbar\omega)
\Biggr\} 
\ .
\end{eqnarray}
The function $\zeta_0(\omega)$ in the denominators is easily calculated as
\begin{eqnarray}\label{zeta0}
\zeta_0(\omega) &=& 
\hbar^2 v^4\kappa 
\Bigg|
{\rm Im}\xi_-(\omega)
\left\{ 
\omega+
\frac{2\overline{\Delta}}{\hbar v}{\rm Re}\eta_-(\omega) 
+\frac{\omega}{v^2}\big|\eta_-(\omega)\big|^2
\right\}
\\
&&\mskip 70 mu + 
{\rm Im}\eta_-(\omega)
\left\{ 
\omega+
\frac{2\overline{\Delta}}{\hbar v}{\rm Re}\xi_-(\omega) 
+\frac{\omega}{v^2}\big|\xi_-(\omega)\big|^2
\right\}\Bigg|\ .
\end{eqnarray}

\subsection{Electric current}

As easily seen, the electric current at $x$ in the TLM chain is given by
\begin{equation}
J(x)=-ev\sum_\sigma \Psi_\sigma^\dag(x)\sigma_y\Psi_\sigma(x) \ ,
\end{equation}
and its NESS average by 
\begin{equation*}
\overline{J}=\left<J(x) \right>_\infty
=-\frac{4e v^2}{\pi}\int_{-\omega_c}^{\omega_c}d\omega
(\hbar \kappa v)^2
{
{\rm Im}\eta_-(\omega) {\rm Im}\xi_-(\omega)
\over |v^2 D(\omega)\Lambda_-(\omega)|^2}
\Big( f_L(\hbar\omega)-f_R(\hbar\omega) \Big) \ .
\label{cur}
\end{equation*}
As in the case of the self-consistent equation, for large $\ell$,
the averaged current reduces to
\begin{equation}
\overline{J}
=-\frac{4e v^2}{\pi}\int\limits_{{|{\overline\Delta}|\over\hbar}<|\omega|<\omega_c}d\omega
(\hbar \kappa v)^2
{
{\rm Im}\eta_-(\omega) {\rm Im}\xi_-(\omega)
\over \zeta_0(\omega)}
\Big( f_L(\hbar\omega)-f_R(\hbar\omega) \Big) \ .
\label{current}
\end{equation}

\subsection{Stability}

Since no general thermodynamic criterion is available for discussing the stabilities of NESS, 
we study the phase stability based on the linear stability of the adiabatic evolution
equation for a spatially uniform order parameter.

Within the adiabatic approximation, the force on the order parameter from the electrons is given
by $\pi \hbar v \lambda \sum_{\sigma}\langle
\Psi_\sigma^\dag(x) \sigma_x \Psi_\sigma(x)\rangle_\infty$, where the order parameter is replaced
with its instantaneous value ${\widetilde\Delta}(t)$.
Then, we have
\begin{eqnarray}
\frac{\partial^2 {\widetilde\Delta}(t)}{\partial t^2}&=&-\omega_0^2
\Big\{
1+ \lambda \
S\big({\widetilde\Delta}(t),V,T_L,T_R\big)
\Big\}
{\widetilde\Delta}(t) \ ,
\end{eqnarray}
where the function $S({\widetilde\Delta},V,T_L,T_R)$ is defined by (\ref{SelfUni2}).

First, we consider the stability of the normal phase where ${\overline\Delta}=0$.
As the linearized equation for ${\widetilde\Delta}(t)$ is given by
$$
\frac{\partial^2 {\widetilde\Delta}(t)}{\partial t^2}=-\omega_0^2
\Big\{
1+ \lambda \
S\big(0,V,T_L,T_R\big)
\Big\}
{\widetilde\Delta}(t)
\ ,
$$
the phase is stable when 
\begin{equation}\label{NormalStable}
\chi_N\equiv
1+ \lambda \
S\big(0,V,T_L,T_R\big)>0 \ .
\end{equation}

Next, we investigate the stability of the phase with 
nonvanishing ${\overline\Delta}$.
Then, two cases should be distinguished: the constant-bias-voltage and 
constant-current cases.
In the former, the linearized equation for $\delta{\widetilde\Delta}(t)
={\widetilde\Delta}(t)-{\overline\Delta}$ is
$$
\frac{\partial^2 {\widetilde\Delta}(t)}{\partial t^2}=-\omega_0^2\lambda
{\overline\Delta}\Bigg(
{\partial S \over \partial {\overline\Delta}}
\Bigg)_V
\delta{\widetilde\Delta}(t)
\ ,
$$
and the phase is stable when 
\begin{equation}\label{ConstantVoltageStable}
\chi_V({\overline \Delta})=\lambda
{\overline\Delta}\Bigg(
{\partial S\over \partial {\overline\Delta}}
\Bigg)_V>0 \ .
\end{equation}
In the latter case, the linearized equation is
$$
\frac{\partial^2 {\widetilde\Delta}(t)}{\partial t^2}=-\omega_0^2\lambda
{\overline\Delta}\Bigg(
{\partial S\over \partial {\overline\Delta}}
\Bigg)_{\overline{J}}
\delta{\widetilde\Delta}(t)
\ ,
$$
and the phase is stable when 
\begin{eqnarray}\label{ConstantCurrentStable}
\chi_I({\overline \Delta})&=& \lambda
{\overline\Delta}\Bigg(
{\partial S\over \partial {\overline\Delta}}
\Bigg)_{\overline{J}}\nonumber\\
&=&\lambda{\overline\Delta}
\Bigg\{
\Bigg(
{\partial S\over \partial {\overline\Delta}}
\Bigg)_V - \Bigg(
{\partial S\over \partial V}
\Bigg)_{\overline\Delta}
\Bigg(
{\partial \overline{J}\over \partial {\overline\Delta}}
\Bigg)_V \bigg/
\Bigg(
{\partial \overline{J}\over \partial V}
\Bigg)_{\overline\Delta}
\Bigg\}>0
\ .
\end{eqnarray}
As will be shown in Appendix~\ref{ap.sta}, the phase with a nontrivial order
parameter at constant current is more stable than that at constant bias voltage.

\section{Nonequilibrium Phase Transitions}

\subsection{Basic formula}

In this section, we study the nonequilibrium phase transitions when the TLM chain couples
symmetrically with two identical reservoirs at temperature $T$: $|v_{\bf k}|=|w_{\bf k}|$,
$\omega_{kL}=\omega_{kR}$, and $T_L=T_R=T$. Since $\eta_-(\omega)=\xi_-(\omega)$ and
$$
{
\omega+{2{\overline\Delta}\over \hbar v}{\rm Re}\eta_-(\omega)
+{\omega\over v^2}|\eta_-(\omega)|^2
\over
\Big|\omega+{2{\overline\Delta}\over \hbar v}{\rm Re}\eta_-(\omega)
+{\omega\over v^2}|\eta_-(\omega)|^2\Big|}={\rm sgn}(\omega)
$$
for $|\hbar\omega|>|{\overline\Delta}|$ 
where sgn$(\omega)$ the sign of $\omega$, we have
\begin{eqnarray*}
&&
S({\overline\Delta},V,T_L,T_R)\Big|_{T_L=T_R=T}
\equiv
S({\overline\Delta},V,T)
=
\int\limits_{{|{\overline\Delta}|\over\hbar}<|\omega|<\omega_c} {d\omega \over\kappa v}
{\rm sgn}(\omega) \big\{f_L(\hbar\omega)+f_R(\hbar\omega)\big\}
\ .
\end{eqnarray*}
Thus, the self-consistent equation (\ref{SelfUni1b}) becomes
\begin{eqnarray}
{1\over \lambda}=-S({\overline\Delta},V,T)
=\int^{\hbar\omega_c}_{|{\overline\Delta}|}
\frac{d \epsilon}{\sqrt{\epsilon^2-{\overline\Delta}^2}}\
\frac{ 2 \sinh(\epsilon/T)}
{\cosh(\frac{eV}{2T})+\cosh(\epsilon/T)}
\ .
\label{SelfUniNum}
\end{eqnarray}
Since the current is carried by electrons having energies near the Fermi energies, 
we further approximate $\xi_-(\omega)=\eta_-(\omega)=i w_0$ and we have
\begin{eqnarray}
\overline{J}
&=&
-\frac{2e vw_0}{\pi(v^2+w_0^2)}\int\limits_{{|{\overline\Delta}|\over\hbar}<|\omega|<\omega_c}d\omega
{
\kappa v
\over |\omega|}
\Big( f_L(\hbar\omega)-f_R(\hbar\omega) \Big) 
\nonumber
\\
&=&
\frac{2G_0}{e}
\int^{\hbar\omega_c}_{|{\overline\Delta}|} d\epsilon 
\frac{\sqrt{\epsilon^2- {\overline\Delta}^2}}
{\epsilon}
\frac{\sinh({eV\over 2T})}{\cosh({eV\over 2T})+\cosh({\epsilon\over T})}
\ ,
\label{currentNum}
\end{eqnarray}
where the normal-state conductance $G_0$ is given by 
\begin{equation}
G_0={e^2 \over \pi\hbar}{2 v w_0\over v^2+w_0^2}
\ .
\end{equation}
Indeed, 
if ${\overline\Delta}=0$ and terms on the order of
$\exp\{-(2\hbar\omega_c-e|V|)/(2T)\}$ are neglected,
one can easily evaluate the integral in 
(\ref{currentNum}), yielding
\begin{eqnarray*}
\overline{J}
=G_0 V \ ,
\end{eqnarray*}
irrespective of the temperature $T$. 
In the rest of this section, the phase transition and nonlinear conduction will be discussed
based on (\ref{SelfUniNum}) and (\ref{currentNum}).

\subsection{Phases at absolute zero temperature \label{subsec. Zero}}

In this subsection, phases at absolute zero temperature are investigated.
Thanks to the formula 
$$
\lim_{T\to 0}{\sinh(y/T)\over \cosh(y/T)+\cosh(x/T)}={\rm sgn}(y) \theta(|y|-|x|)
\  
$$
with the step function $\theta$,
the function $S({\overline\Delta},V,T)$ and the current at $T=0$ are given by
\begin{eqnarray}
S({\overline\Delta},V,0)=-2 
\int^{\hbar\omega_c}_{|{\overline\Delta}|}
\frac{d \epsilon}{\sqrt{\epsilon^2-{\overline\Delta}^2}}\
\theta\Big(\epsilon -\Big|\frac{eV}{2}\Big|\Big)
\ ,
\label{SelfUniNum2}
\end{eqnarray}
\begin{eqnarray}
\overline{J}
&=&
\frac{2G_0}{e}{\rm sgn}(V)
\int^{\hbar\omega_c}_{|{\overline\Delta}|} d\epsilon 
\frac{\sqrt{\epsilon^2- {\overline\Delta}^2}}{\epsilon}
\theta\Big( \Big|\frac{eV}{2}\Big|-\epsilon \Big)
\ .
\label{currentNum2}
\end{eqnarray}

The stability index of the normal phase ${\overline \Delta}=0$ is, then,
\begin{equation}
\chi_N=1+\lambda S(0,V,0)=2\lambda \ln{|V|\over V_{10}} \ ,
\end{equation}
where $V_{10}$ is defined by $V_{10}={2\hbar\omega_c\over e}\exp(-{1\over 2\lambda})$. Hence, the normal phase
is stable if $|V|>V_{10}$ and unstable if $|V|<V_{10}$.
As mentioned in the previous subsection, the average current is given by
\begin{equation}
\overline{J}=G_0 V \ .
\end{equation}

For $|{\overline \Delta}|\ge |eV|/2$, (\ref{SelfUniNum2}) reduces to
\begin{equation}
S({\overline\Delta},V,0)=
-2 \int^{\hbar\omega_c}_{|{\overline\Delta}|}
\frac{d \epsilon}{\sqrt{\epsilon^2-{\overline\Delta}^2}}\
=-2\cosh^{-1}{\hbar \omega_c\over |{\overline\Delta}|}
\ ,
\end{equation}
and (\ref{SelfUniNum}) has a nontrivial solution,
\begin{equation}
|{\overline\Delta}|={\hbar \omega_c 
\over\cosh{1\over 2\lambda}}
\equiv \Delta_0
\ ,
\end{equation}
irrespective of the bias voltage $|V|\le V_{20}\equiv 2\Delta_0/e$.
Its stability index at constant bias voltage is always positive:
\begin{equation}
\chi_V=\lambda{\overline\Delta}\Big({\partial S\over \partial {\overline\Delta}}
\Big)_V={2\lambda\hbar\omega_c\over \sqrt{(\hbar\omega_c)^2-|{\overline\Delta}|^2}}
>0 \ .
\end{equation}
In short, the ordered phase exists for $|V|\le V_{20}$ and
is stable at constant bias voltage.
According to the inequality $\chi_I>\chi_V$ shown in Appendix~\ref{ap.sta}, this phase is also stable at constant
current.
Since the Fermi energies of the two reservoirs fall into the energy gap, the phase is
insulating:
\begin{equation}
\overline{J}= 0 \ .
\end{equation}

Because of $V_{20}/V_{10}=2/(1+e^{-1/\lambda})>1$, 
there might be a first-order phase
transition between the normal and insulating ($|{\overline\Delta}|=\Delta_0$) phases
for $V_{10}<|V|<V_{20}$.
This suggests the existence of another solution $|{\overline\Delta}|$ of (\ref{SelfUniNum}) 
satisfying $0<|{\overline\Delta}|<\Delta_0$.  
Indeed, for $|{\overline \Delta}|<|eV|/2$, 
\begin{equation}
S({\overline\Delta},V,0)=
-2 \int^{\hbar\omega_c}_{|eV|/2}
\frac{d \epsilon}{\sqrt{\epsilon^2-{\overline\Delta}^2}}\
=-2\Big\{
\cosh^{-1}{\hbar \omega_c\over |{\overline\Delta}|}
-\cosh^{-1}{|eV|\over 2|{\overline\Delta}|}\Big\}
\ ,
\end{equation}
and thus
(\ref{SelfUniNum}) has a nontrivial solution for $V_{10}\le |V|\le V_{20}$:
\begin{equation}\label{ZeroPhase2}
|{\overline\Delta}|=
\Delta_0
\sqrt{{|V|-V_{10}\over V_{20}-V_{10}}\ \bigg\{{|V|\over V_{20}}
+{V_{10}(V_{20}-|V|)\over V_{20}(V_{20}-V_{10})}
\bigg\}
} \equiv \Delta_1(\lambda,V)\ 
(<\Delta_0)
\ .
\end{equation}
This phase is unstable at constant bias voltage since the stability index
is negative for $|{\overline\Delta}|<|eV|/2<\hbar\omega_c$:
\begin{equation}
\chi_V=\lambda{\overline\Delta}\Big({\partial S\over \partial {\overline\Delta}}
\Big)_V=
{2\lambda\hbar\omega_c\over \sqrt{(\hbar\omega_c)^2-|{\overline\Delta}|^2}}
-{\lambda|eV|\over \sqrt{(eV)^2/4-|{\overline\Delta}|^2}}
<0 \ .
\end{equation}
In contrast, {\it at constant current}, this phase is {\it stable} because 
the stability index is positive:
$\chi_I>0$
(for a proof, see Appendix~\ref{ap.sta}). In this case, the average current is given by
\begin{eqnarray}
&&\overline{J}
=
\frac{2G_0}{e}{\rm sgn}(V)
\int^{|eV|/2}_{|{\overline\Delta}|} d\epsilon 
\frac{\sqrt{\epsilon^2- {\overline\Delta}^2}}{\epsilon}
\nonumber\\
&&~~=\frac{2G_0}{e}|{\overline\Delta}|{\rm sgn}(V)
\Big\{
\sqrt{\Big({eV\over 2{\overline\Delta}}\Big)^2-1}
-\tan^{-1}\sqrt{\Big({eV\over 2{\overline\Delta}}\Big)^2-1}
\Big\}\equiv J_2(\lambda,V)
\ .
\end{eqnarray}

\begin{figure}[b]
\parbox{\halftext}{
\rotatebox[origin=c]{-90}{\includegraphics[scale=0.26]
	{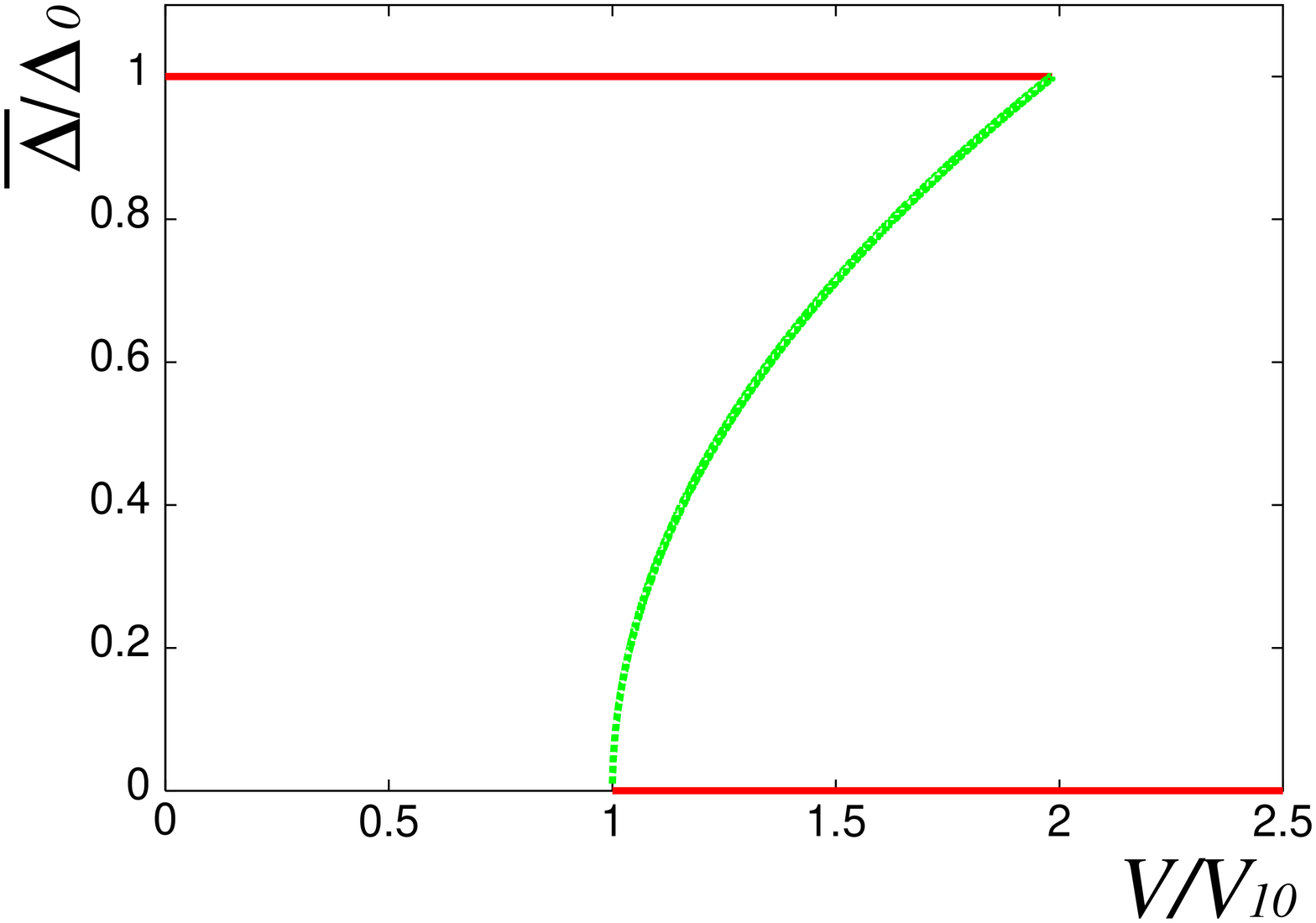}}
	\caption{Bias-voltage dependence of the order parameter at absolute zero temperature.
In the constant-bias-voltage case, the solid lines correspond to the stable phases 
and the dashed curve to the unstable phase.
All phases are stable in the constant-current case.}
	\label{vol-gap-Zero}}
\hfill
\parbox{\halftext}{\vspace{-11.5mm}
\rotatebox[origin=c]{-90}{\includegraphics[scale=0.26]
	{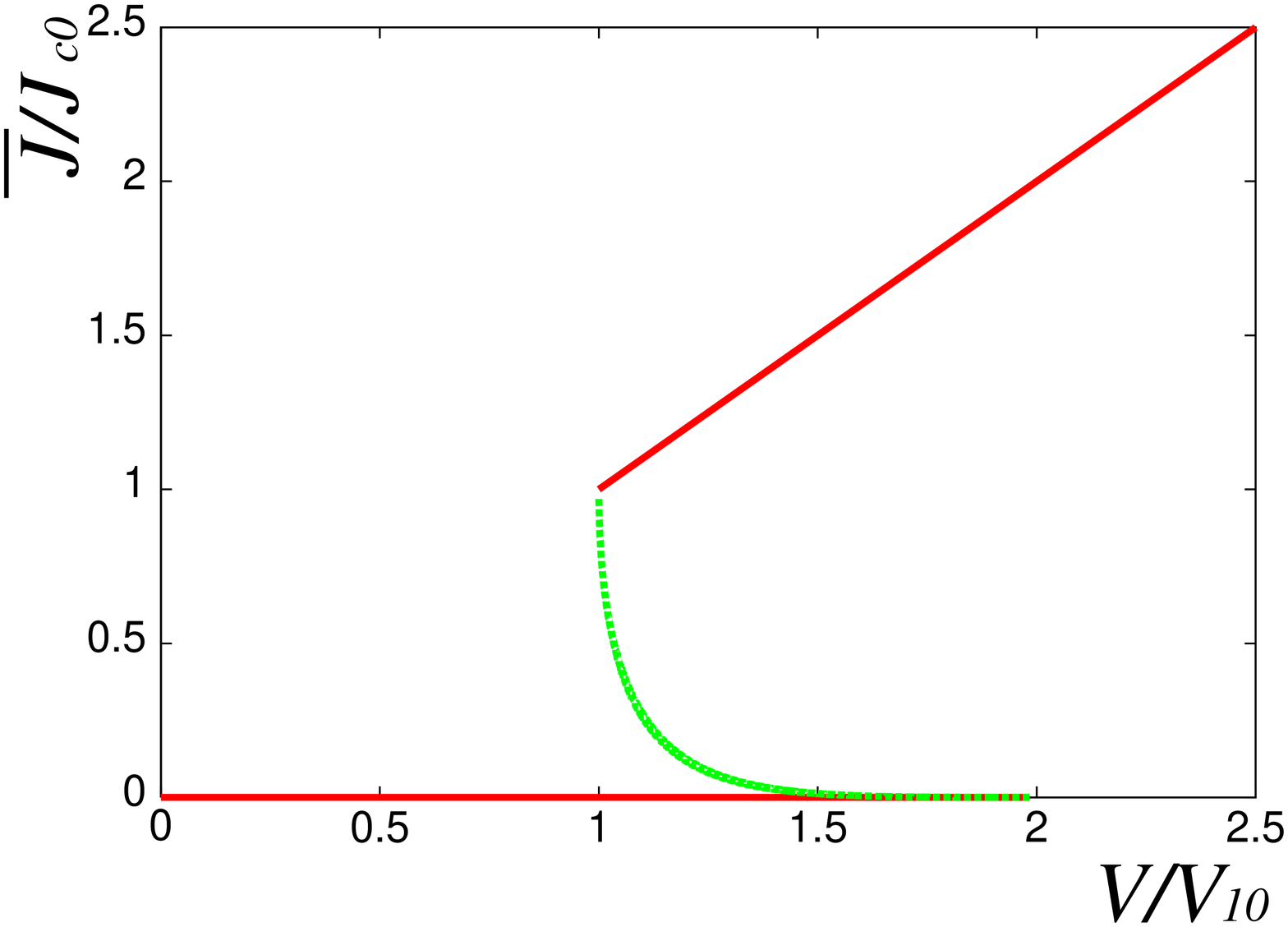}}
	\caption{Current versus bias voltage at absolute zero temperature. 
The solid lines and dashed curve correspond 
to those of Fig.~\ref{vol-gap-Zero}.}
	\label{vi@Zero}}
\end{figure}

\begin{figure}[t]
\parbox{\halftext}{
\rotatebox[origin=c]{-90}{\includegraphics[scale=0.26]
	{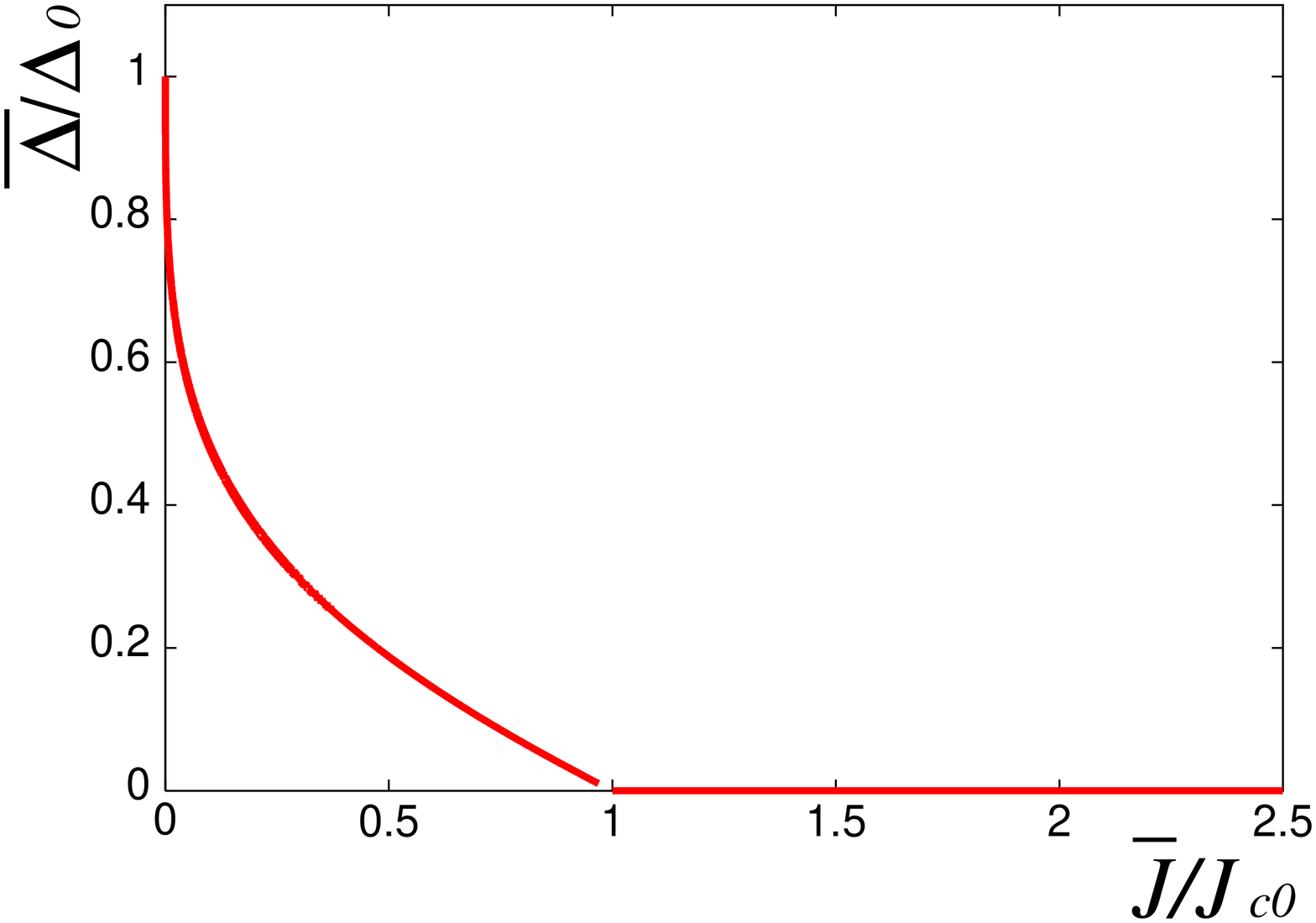}}
	\caption{Current dependence of the order parameter at absolute zero temperature.
In the constant-current case, all phases are stable. Note that the point
$(\overline{J},|{\overline\Delta}|)=(0,\Delta_0)$ 
corresponds to the insulating phases for $V_{10}\le |V|\le V_{20}$.}
	\label{I-gap-Zero}}
\hfill
\parbox{\halftext}{\vspace{-8.9mm}
\rotatebox[origin=c]{-90}{\includegraphics[scale=0.26]
	{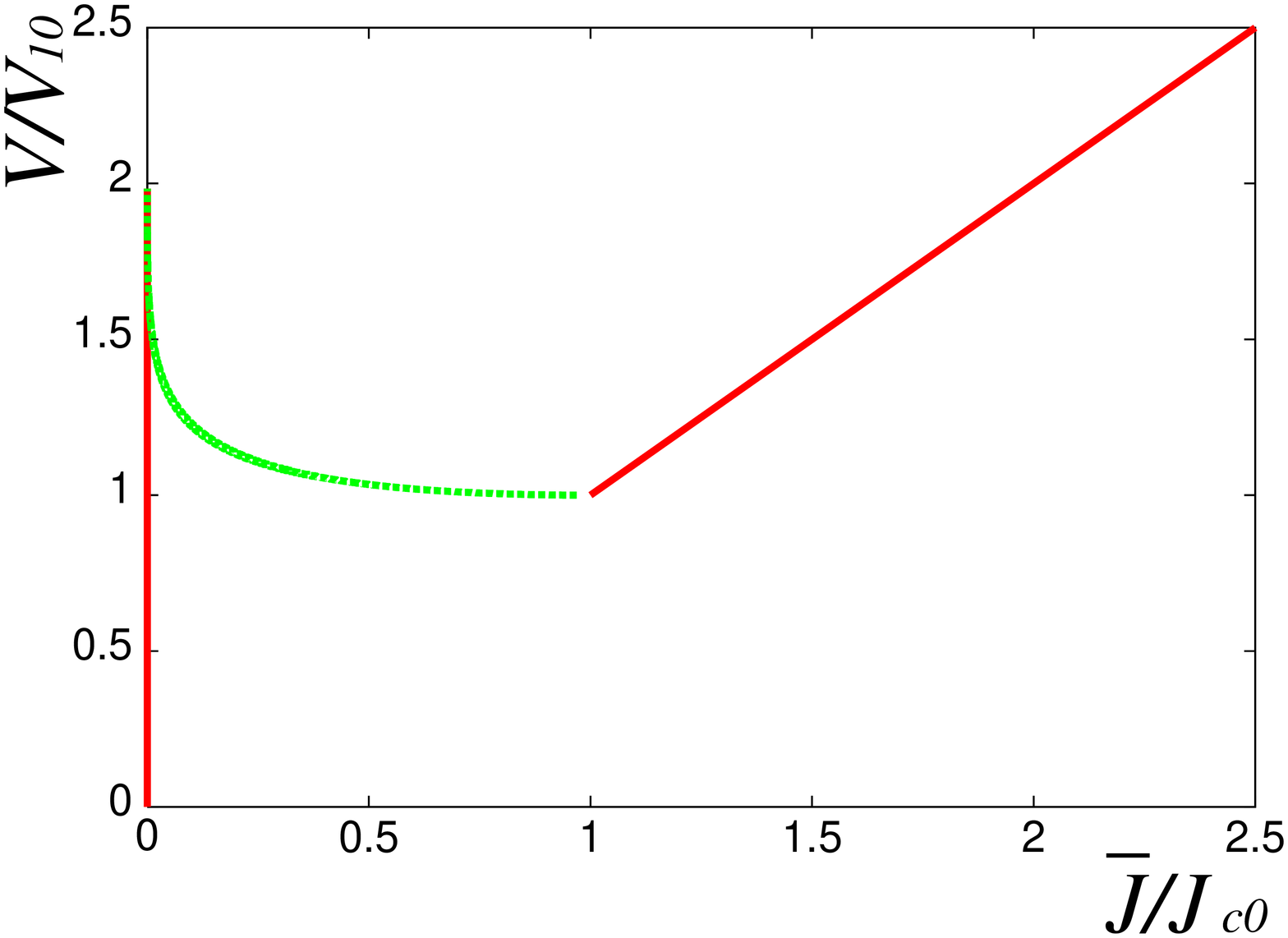}}
	\caption{Bias voltage versus current at absolute zero temperature (constant current). 
The curve is essentially the same as Fig.~\ref{vi@Zero}. Only stabilities are different.}
	\label{iv@Zero}}
\end{figure}

These results are summarized in Figs.~\ref{vol-gap-Zero}-\ref{iv@Zero}.
At constant bias voltage (cf. Fig.~\ref{vol-gap-Zero}), the insulating phase is stable up to the first threshold 
voltage
$eV_{10}$, which is almost half of the zero-bias gap $2\Delta_0$ for small $\lambda$.
Beyond the second threshold voltage $eV_{20}$ which is equal to the zero-bias gap $2\Delta_0$, 
only the normal phase is stable. 
Between the two threshold voltages, i.e. $V_{10}\le |V|\le V_{20}$, both the insulating and normal phases are stable, 
and a first-order phase transition between them is possible.
Moreover, there exists an unstable phase separating the two stable ones, as shown by the dashed curve 
(i.e., the curve satisfying $|{\overline\Delta}|<e|V|/2$)
in Fig.~\ref{vol-gap-Zero}.
The current-voltage characteristics are shown in Fig.~\ref{vi@Zero}.
Note that
the first-order transition corresponds to a sudden change in the current.
The three regions discussed above are summarized in Table~\ref{table for zero}.
\begin{table}[b]
\caption{Three regions at absolute zero temperature (constant-bias-voltage case)}\vskip 3pt
\label{table for zero}
\begin{center}
\begin{tabular}{c||c|c|c}
\hline
& $V$ & $|{\overline\Delta}|$ & $\overline{J}$
\\
\hline\hline
region~A & $0\le V \le V_{10}$ & $ \Delta_0(\lambda)$&
0 
\\
\hline
region~B & $V_{10}\le V \le V_{20}$ & $0,\ \Delta_0(\lambda),\ 
\Delta_1(\lambda,V)$ &
$G_0 V$,\ 0,\ $J_2(\lambda,V)$
\\
\hline
region~C & $V_{20}< V$ & 0 & $G_0 V$\\
\hline
\end{tabular}
\end{center}
\end{table}

On the other hand, 
at constant current (see Fig.~\ref{I-gap-Zero}), the unstable phase mentioned above is
stabilized, and when one increases the current, a second-order phase transition to the normal phase 
occurs at the critical current $|{\overline J}|=J_{c0}\equiv G_0V_{10}$. 
Near the critical current, the order parameter 
changes linearly with respect to the current:
\begin{equation}
|{\overline\Delta}|\simeq {2\Delta_0\over\pi}{J_{c0}-|\overline{J}|\over G_0V_{20}} \ .
\label{GapCurrentZero}
\end{equation}
The corresponding voltage-current characteristics are shown in Fig.~\ref{iv@Zero};
a region $0<\overline{J} <J_{c0}$ with negative differential conductivity appears.

The existence of negative differential conductivity can be understood as follows.
When no bias voltage is applied, the system is in an insulating phase with an energy gap $2\Delta_0$, which
is robust against low bias voltages. 
When the current starts to flow, electrons with energy larger than the gap should exist, and
the bias voltage should be on the order of the zero-bias gap: $|V|\sim 2\Delta_0/e$.
At the same time, the gap is reduced by the existence of the current.
As the current increases, the gap $2|{\overline\Delta}|$ is reduced further, and the corresponding bias voltage
$|V|\sim 2|{\overline\Delta}|/e$ becomes smaller. Thus, negative differential conductivity does appear. 
When the current reaches the critical current, the gap disappears and the phase becomes normal.
As the normal-phase conductivity is positive, negative differential conductivity appears only up to
the critical current: $|\overline{J}|<J_{c0}$.

\subsection{Phases at finite temperature at constant bias voltage \label{subsec. Finite}}

In this subsection, we investigate the finite-temperature phases at constant bias voltage. 

\noindent
\underline{(A)\ Phase Diagram}

\begin{wrapfigure}{r}{7cm}
\rotatebox[origin=c]{-90}{
\includegraphics[width=5cm]{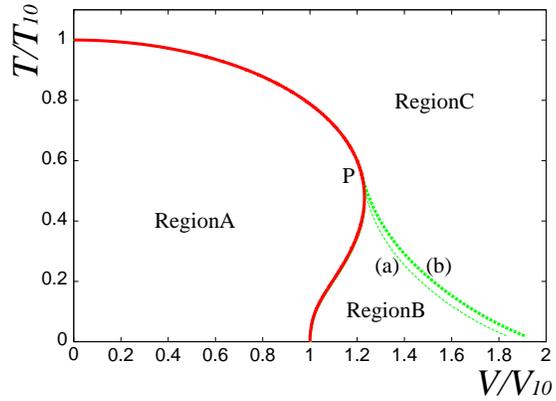}
}
\caption{Phase diagram at constant bias voltage. A unique stable ordered phase occurs in region~A, 
one stable ordered phase and the normal phase in region~B and only the normal phase in region~C. 
The solid curve represents the second-order phase transition, and the dashed curves represent the 
first-order phase transition. For comparison, the phase 
transition curve for (a)~$\lambda=4.8$ is shown together with that for (b) $\lambda=3.0$.}
\label{PhaseDiagram}
\end{wrapfigure}

Let us begin with the investigation of the phase diagram.
As in the zero-temperature case, three regions exist in the $VT$-plane; 
one with a unique stable ordered phase (region A), one where ordered and normal phases are stable 
(region B) and one with the normal phase (region C). 
The three regions are depicted in Fig.~5.
The boundary curve between region~A and the 
others is implicitly given by
\begin{eqnarray}
&&0={\chi_N\over 2\lambda}={1\over 2\lambda}+{1\over 2}S(0,V,T)
={1\over 2\lambda}-
\int^{\hbar\omega_c/T}_0
\frac{d \epsilon}{\epsilon}\
\frac{ \sinh\epsilon}
{\cosh(\frac{eV}{2T})+\cosh\epsilon}
\ ,
\nonumber
\end{eqnarray}
where $\chi_N$ is the stability index of the normal phase.
When terms on the order of $\exp\{-(2\hbar\omega_c-e|V|)/(2T)\}$
are neglected,
it reduces to
\begin{eqnarray}
&&\log{eV_{10}\over 2T}=\int_0^\infty d\epsilon \ \log \epsilon \frac{\cosh(\frac{eV}{2T}) \cosh \epsilon +1}
{(\cosh(\frac{eV}{2T})+\cosh \epsilon)^2}
\equiv\phi\Big({eV\over 2T}\Big)
\ ,
\label{scaling}
\end{eqnarray}
where the function $\phi$ is defined by the integral of the middle term.
When $V=0$, (\ref{scaling}) leads to
\begin{eqnarray}
\log{eV_{10}\over 2T}=\int_0^\infty d\epsilon \ \frac{\log \epsilon }
{1+\cosh \epsilon}=-\log{2e^\gamma\over\pi}
\quad {\rm or}
\quad T=T_{c0}\equiv {e^\gamma\over\pi}eV_{10} 
\ ,
\end{eqnarray}
where $\gamma$ is the Euler constant and $T_{c0}$ 
corresponds to the transition temperature at zero bias voltage. Then, (\ref{scaling}) reads
\begin{eqnarray}
&&\log\Big({\pi\over 2e^\gamma}{T_{c0}\over T}\Big)=\phi\Big({\pi\over 2e^\gamma}{V\over V_{10}}{T_{c0}\over T}\Big)
\ .
\label{scaling2}
\end{eqnarray}
This implies that the boundary between region A and the others is independent of the coupling constant once
it is plotted in terms of $V/V_{10}$ and $T/T_{c0}$ (cf. the solid curve in Fig.~\ref{PhaseDiagram}).
We note that the boundary curve (\ref{scaling2}) can be expressed as $|V|=V_1(T)$ in terms of a single-valued
function $V_1(T)$ of $T$, which will be referred to as the first threshold voltage.

The phase boundary curve (\ref{scaling2}) indicates the bias-induced decrease of the transition 
temperature for low $|V|$ and the temperature-induced increase of the first threshold voltage 
for low $T$. Indeed, for $e|V|\ll T$,
with the aid of the formula $\phi'(0)=0$ and
$$
\phi''(0)=\int_0^\infty d\epsilon \ \frac{\log \epsilon (\cosh \epsilon -2)}
{(1+\cosh \epsilon)^2}={7\zeta(3)\over 2\pi^2}
\ ,
$$ 
(\ref{scaling2}) reduces to
\begin{eqnarray}
T\equiv T_c(V)=T_{c0} \exp\Big[-{7\zeta(3)\over 16e^{2\gamma}}\Big({VT_{c0}\over V_{10}T}\Big)^2\Big]\simeq
T_{c0}\Big\{1-{7\zeta(3)\over 16e^{2\gamma}}\Big({V\over V_{10}}\Big)^2 \Big\}\ ,
\label{TcDec}
\end{eqnarray}
where $\zeta(n)$ is the Riemann zeta function.
Thus, the transition temperature $T_c(V)$ decreases with the bias voltage $V$. 
On the other hand, when $T\ll e|V|$, 
one may apply the standard technique of evaluating the low-temperature properties
of the free fermion gas, and 
we have
\begin{eqnarray}
\phi\Big({eV\over 2T}\Big)
&=&-\int_0^\infty d\epsilon \log\Big({\epsilon\over 2T}\Big){d\over d\epsilon}\Big[
{1\over e^{(\epsilon-eV)/(2T)}+1}+{1\over e^{(\epsilon+eV)/(2T)}+1}
\Big]\nonumber\\
&\simeq& 
\log{e|V|\over 2T}-{2\pi^2\over 3}\Big({T\over eV}\Big)^2
=\log{e|V|\over 2T}-{2e^{2\gamma}\over 3}\Big({TV_{10}\over T_{c0}V}\Big)^2
\ ,
\end{eqnarray}
up to $T^2$.
Thus, the first threshold voltage $V_1(T)$ near absolute zero temperature
is given by
\begin{eqnarray}
|V|=V_1(T)\simeq V_{10}\exp\Big[{2e^{2\gamma}\over 3}\Big({TV_{10}\over T_{c0}V_1}\Big)^2\Big]
\simeq V_{10}\Big\{1+{2e^{2\gamma}\over 3}\Big({T\over T_{c0}}\Big)^2\Big\}
\ ,
\label{temp.scaling}
\end{eqnarray}
which shows that the first threshold voltage increases as temperature increases.
Careful asymptotic analysis indicates that ${T/(e|V|)}\sim 0.05$
is the upper bound  where the estimation~(\ref{temp.scaling}) is valid.

On the other hand,
the boundary curve $|V|=V_2(T)$ between regions B and C (the dashed curve in Fig.~\ref{PhaseDiagram})
is derived by solving the self-consistent equation (\ref{SelfUniNum}) for nonvanishing order parameters. 
It starts from the point 
$(|V|,T)=(V_{20},0)$
and terminates 
at a point P
on the boundary curve $|V|=V_1(T)$
between region A and the others 
(see Fig.~5).
The behaviour near the terminating point P 
can be investigated on the basis of the Ginzburg-Landau expansion of the self-consistent equation (\ref{SelfUniNum}).
Up to ${\overline \Delta}^4$, the self-consistent equation becomes 
\begin{equation}
{K_2\over 2}\Big({{\overline\Delta}\over T}\Big)^2-{K_4\over 8}\Big({{\overline\Delta}\over T}\Big)^4={\chi_N\over 2\lambda}
\ , \label{GLExp}
\end{equation}
where $\chi_N=1+\lambda S(0,V,T)$ is the stability index of the normal phase, and the coefficients $K_2$ and $K_4$ depend
only on the ratio $V/T$ (their concrete expressions
are given in Appendix~\ref{ap.GL}). 
As $V/T$ increases, $K_2$ changes sign 
from minus to plus at $V/T\simeq 2.1865 \times V_{10}/T_{c0}$ and $K_4$ is positive there.
Hence,
when $K_2<0$, the quartic polynomial in the left-hand side of (\ref{GLExp}) has one maximum value 0 
at $|{\overline\Delta}|=0$. 
In this case, when $\chi_N>0$, (\ref{GLExp}) has no solution (region C), and 
when $\chi_N<0$, (\ref{GLExp}) has one nonvanishing solution in $|{\overline\Delta}|$ (region A).
On the other hand, when $K_2>0$, the quartic polynomial in the left-hand side of (\ref{GLExp}) has a
local minimum 0 at $|{\overline\Delta}|=0$, and a local maximum $K_2^2/(2K_4)$ at 
$|{\overline\Delta}|=T \sqrt{2K_2/K_4}$.
Then, with respect to $|{\overline\Delta}|$,
(\ref{GLExp}) has no solution when $K_2^2/(2K_4)<\chi_N/(2\lambda)$ (region C), 
two nonvanishing solutions when $K_2^2/(2K_4)>\chi_N/(2\lambda)>0$ (region B) and one nonvanishing 
solution when $\chi_N/(2\lambda)<0$ (region A).
Therefore, the boundary curve $|V|=V_2(T)$ near point P is given by $K_2^2/(2K_4)=\chi_N/(2\lambda)$.
Then,
since the curve $|V|=V_1(T)$
corresponds to $\chi_N=0$, the simultaneous solution of (\ref{scaling2}) and $K_2=0$ is 
the terminating point P: $T\equiv T^*\simeq 0.5571 \times T_{c0}$ and $|V|\simeq 1.2181\times V_{10}$.
Note that, in contrast to the 
first-threshold-voltage
curve $|V|/V_{10}=V_1(T)/V_{10}$,
the curve $|V|/V_{10}=V_2(T)/V_{10}$ depends not only on $T/T_{c0}$
but also on $\lambda$ (cf. Fig.~\ref{PhaseDiagram}). 
\bigskip
\begin{figure}[t]
\parbox{\halftext}{
	\rotatebox[origin=c]{-90}{\includegraphics[scale=0.26]
	{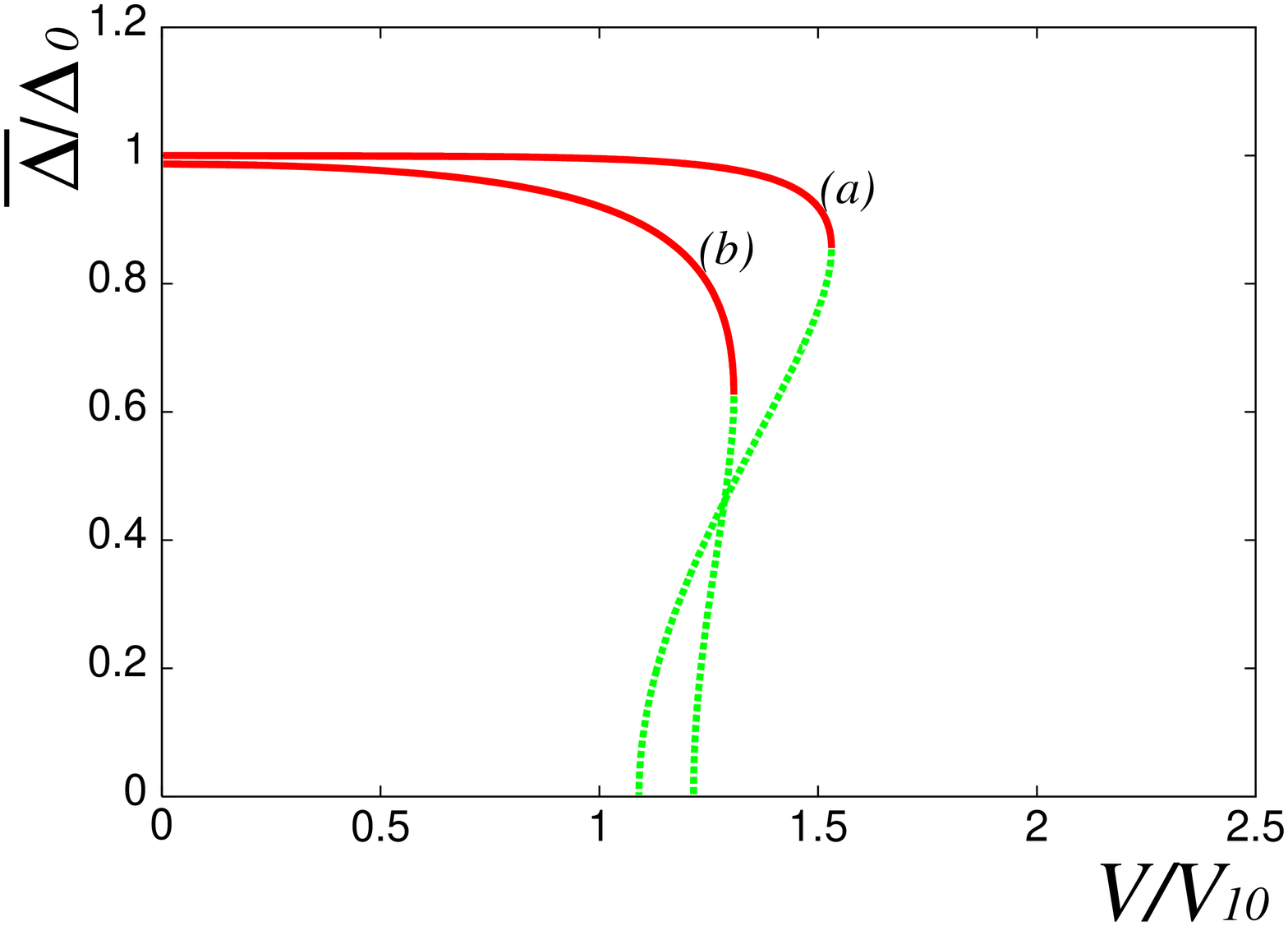}}
	\caption{The order parameter versus bias voltage for 
	(a) $T=0.01\exp(-1/2\lambda)|eV_{10}|$ $(=0.194 T_{c0})$ and 
        (b) $T=0.02\exp(1/2\lambda)|eV_{10}|$ $(=0.389 T_{c0})$.
	The solid lines correspond to the stable phase, and the dashed 
	lines correspond to the unstable phase.}
	\label{vg@low}}
\hfill
\parbox{\halftext}{
	\rotatebox[origin=c]{-90}{\includegraphics[scale=0.26]
	{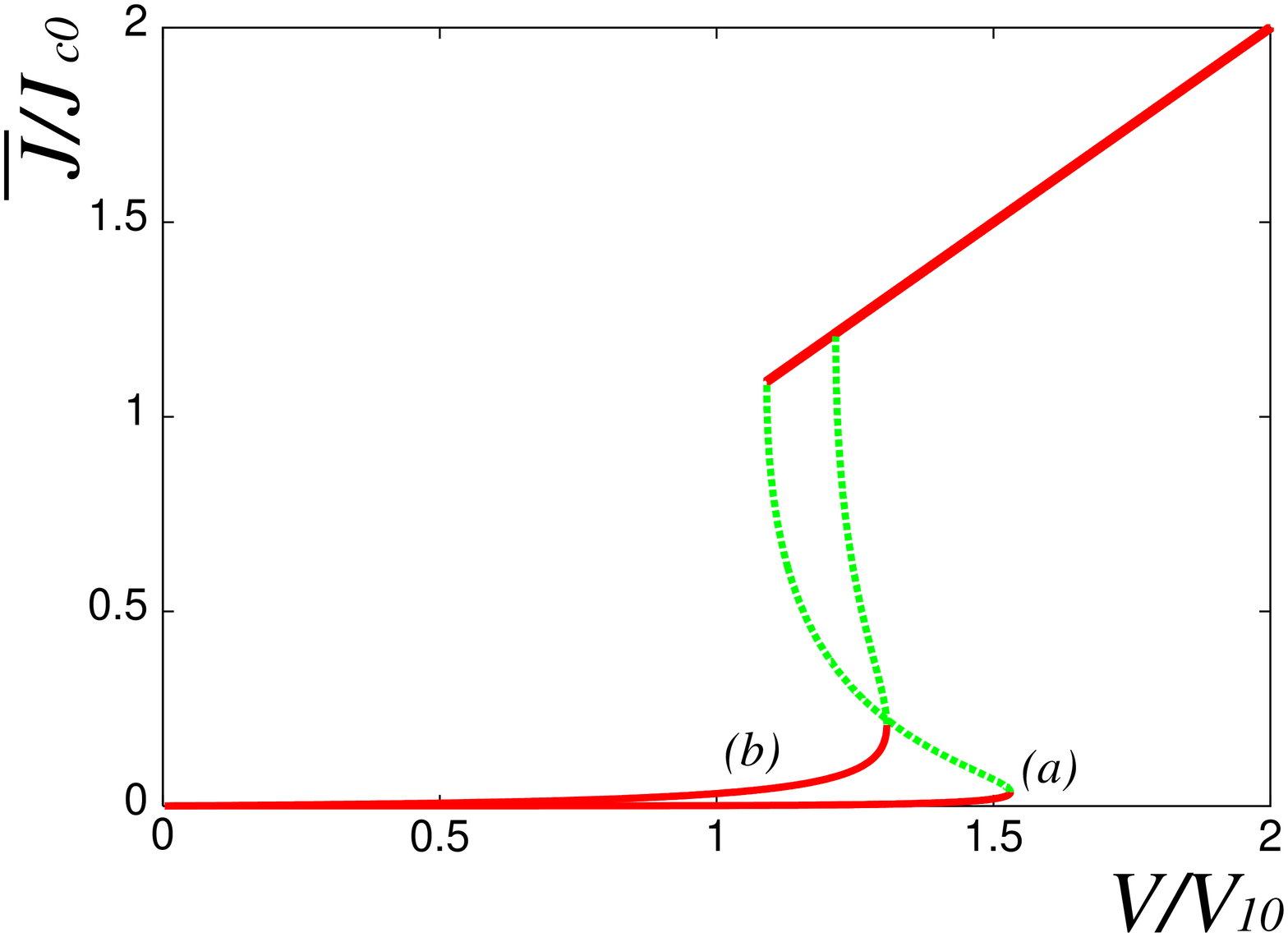}}
	\caption{Current versus bias voltage for 
	(a) $T=0.01\exp(1/2\lambda)|eV_{10}|$ $(=0.194 T_{c0})$ and (b) $T=0.02\exp(1/2\lambda)|eV_{10}|$ $(=0.389 T_{c0})$.
	The solid lines correspond to the stable phase, and the dashed 
	lines correspond to the unstable phase.}
	\label{vi@low}}
\end{figure}
\begin{figure}[t]
\parbox{\halftext}{
	\rotatebox[origin=c]{-90}{\includegraphics[scale=0.26]
	{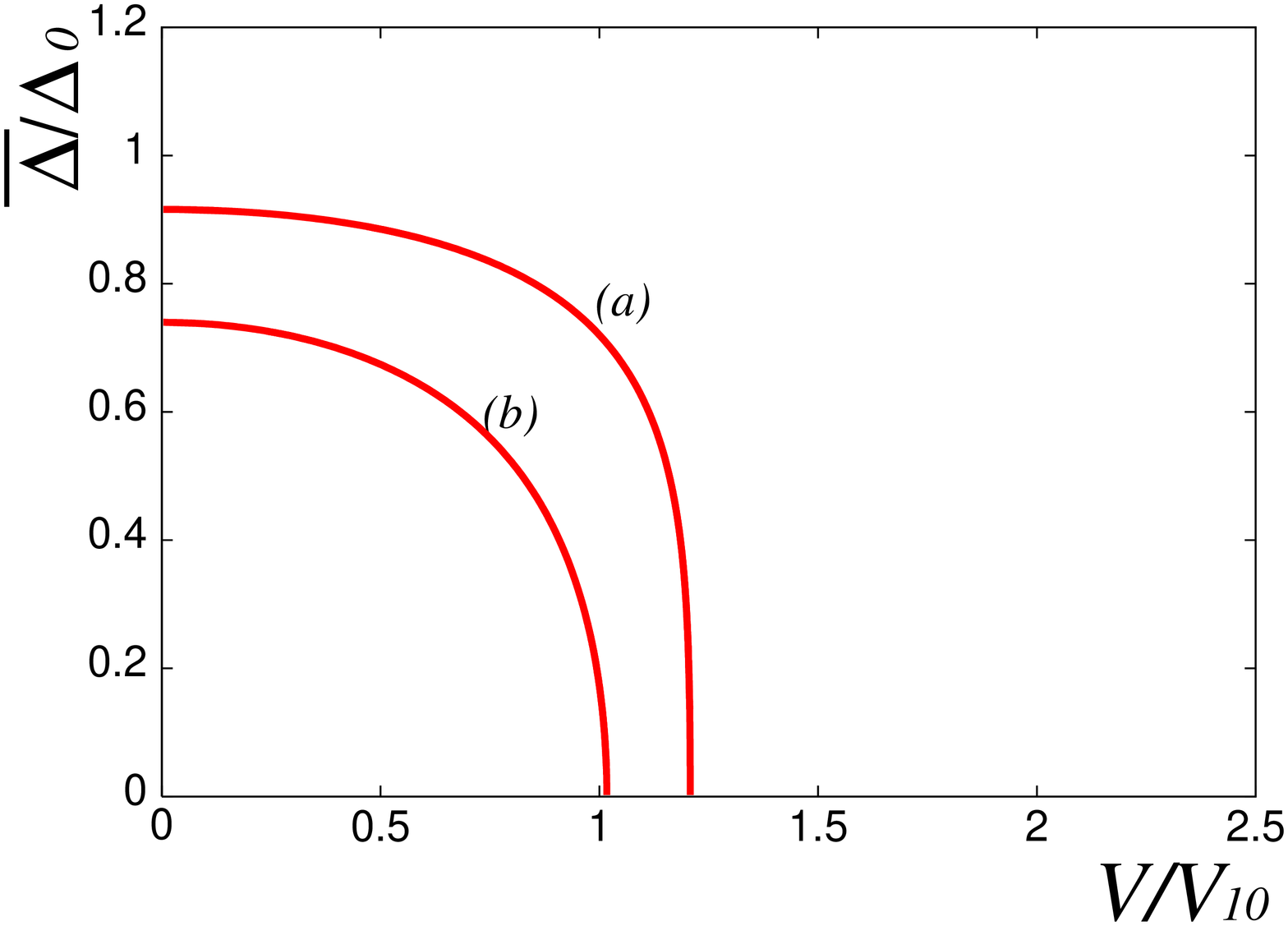}}
	\caption{The order parameter versus bias voltage for 
	(a) $T=0.03\exp(1/2\lambda)|eV_{10}|$ $(=0.583 T_{c0})$ and (b) $T=0.04\exp(1/2\lambda)|eV_{10}|$ $(=0.778 T_{c0}$). 
	All phases shown in this figure are stable.}
	\label{vg@high}}
\hfill
\parbox{\halftext}{\vspace{-4mm}
	\rotatebox[origin=c]{-90}{\includegraphics[scale=0.26]
	{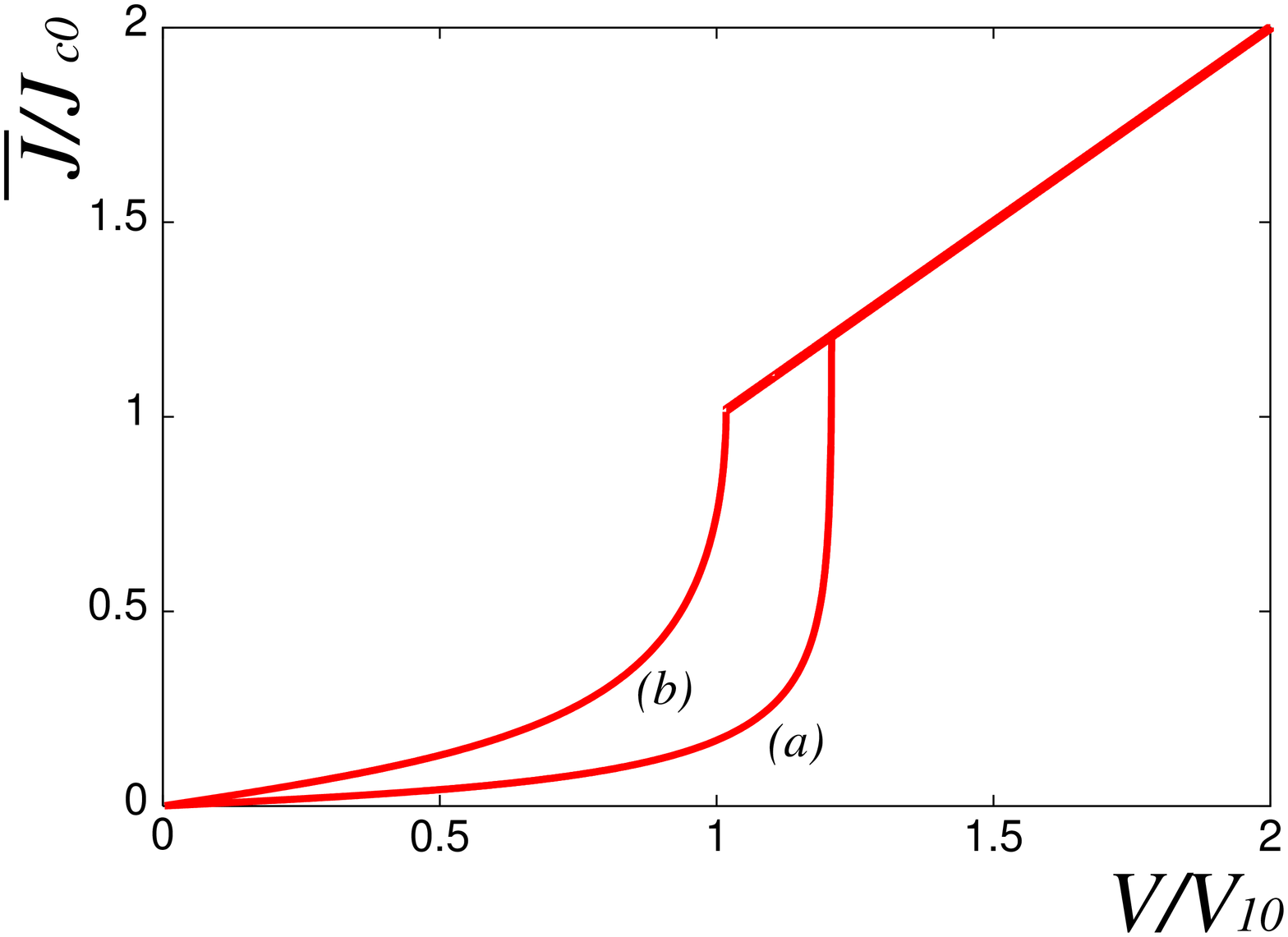}}
	\caption{Current versus bias voltage for 
	(a) $T=0.03\exp(1/2\lambda)|eV_{10}| (=0.583 T_{c0})$ and (b) $T=0.04\exp(1/2\lambda)|eV_{10}| (=0.778 T_{c0})$. 
	All phases shown in this figure are stable.}
	\label{vi@high}}
\end{figure}


\begin{figure}[t]
\vspace{-4mm}
\parbox{\halftext}{\rotatebox[origin=c]{-90}{\includegraphics[scale=0.26]
	{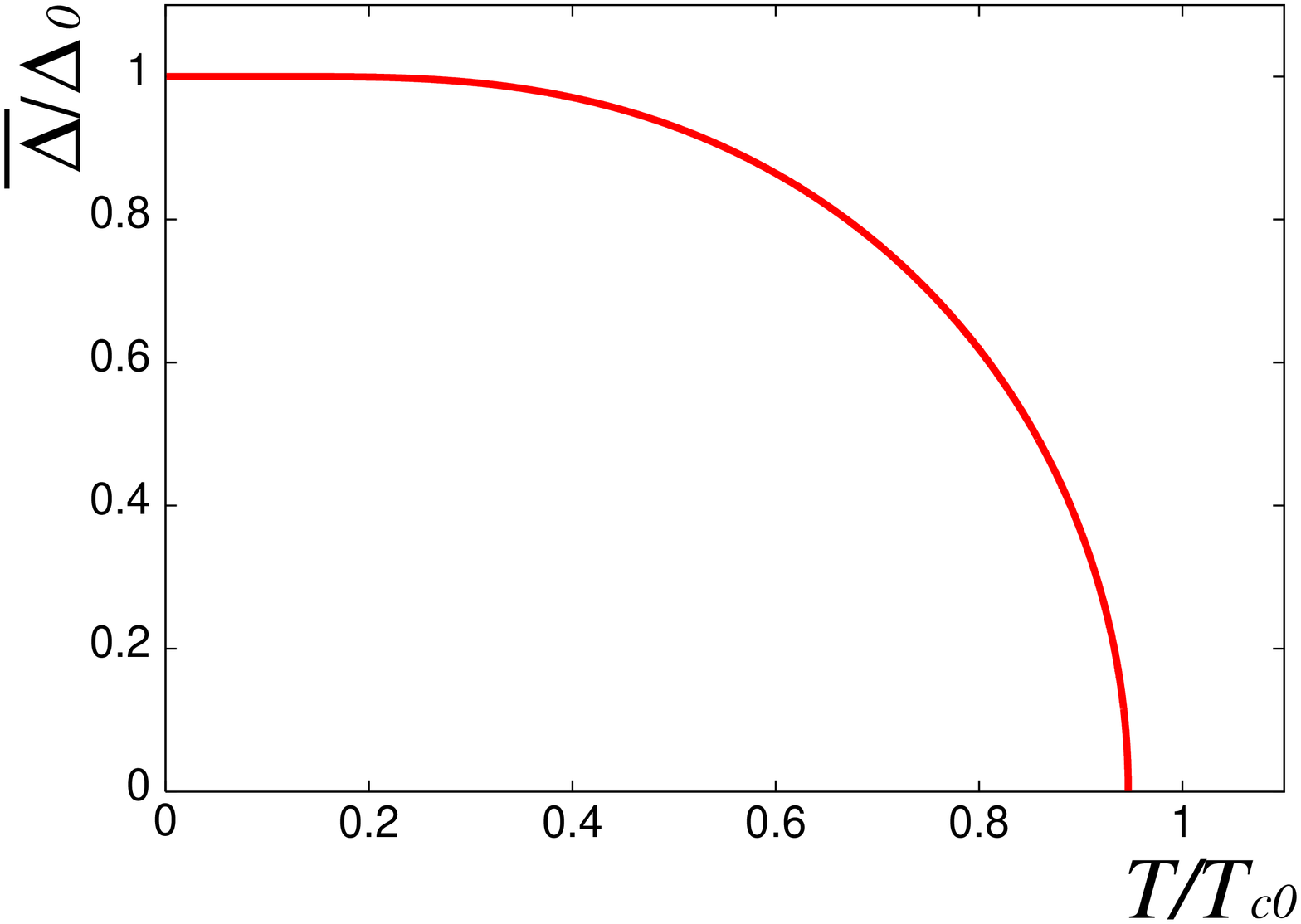}}\vspace{-5mm}
	\caption{The order parameter versus temperature for $V=0.551V_{10}$. The
 	solid lines correspond to the stable phase and the dashed lines correspond 
	to the unstable phase.
}
	\label{GT@0.050}
}
\hfill
\parbox{\halftext}{\rotatebox[origin=c]{-90}{\includegraphics[scale=0.26]
	{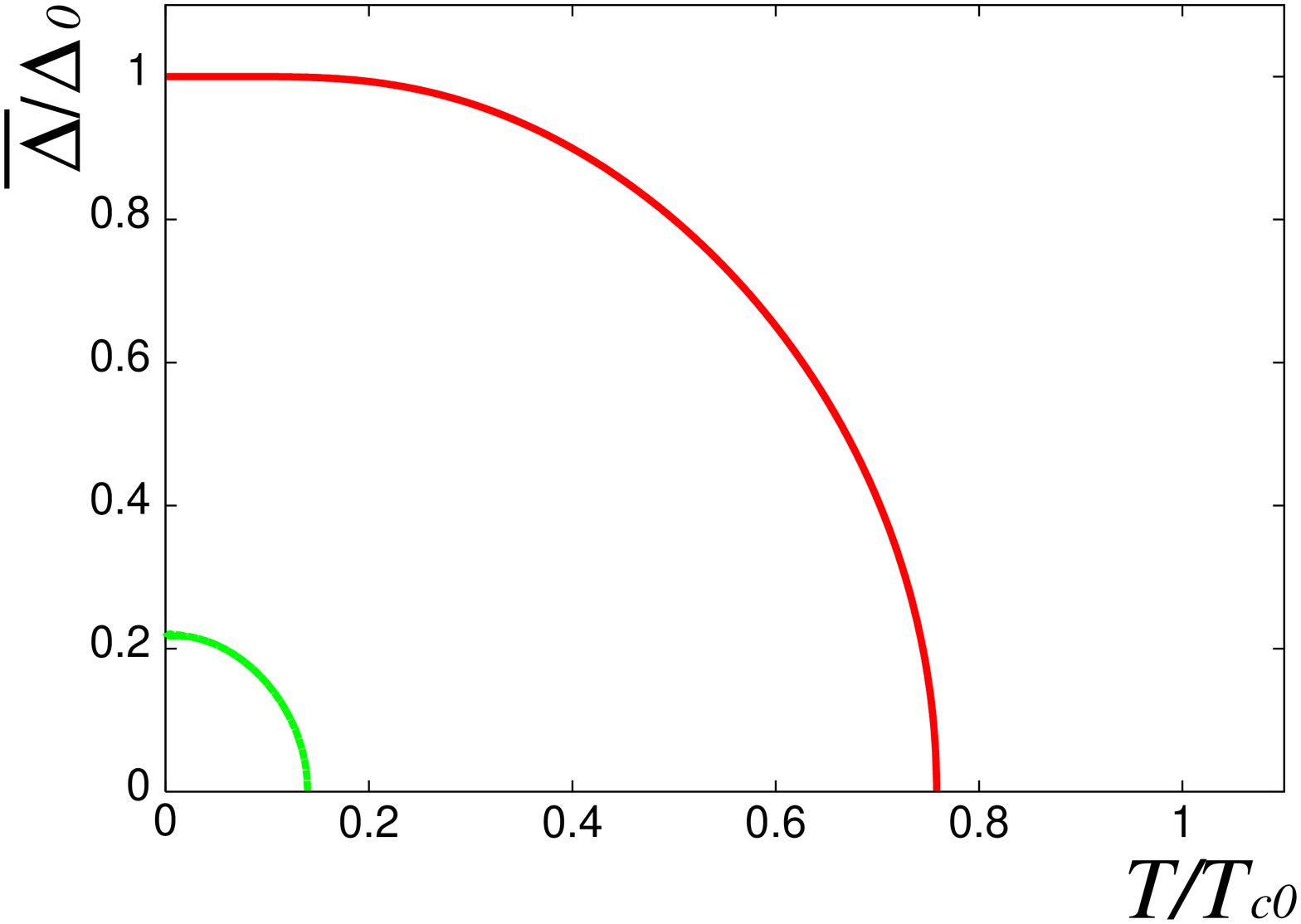}}\vspace{-5mm}
	\caption{The order parameter versus temperature for $V=1.047V_{10}$. The
 	solid lines correspond to the stable phase and the dashed lines correspond 
	to the unstable phase.}
	\label{GT@0.095}}
\end{figure}
\begin{figure}[t]
\parbox{\halftext}{\rotatebox[origin=c]{-90}{\includegraphics[scale=0.26]
	{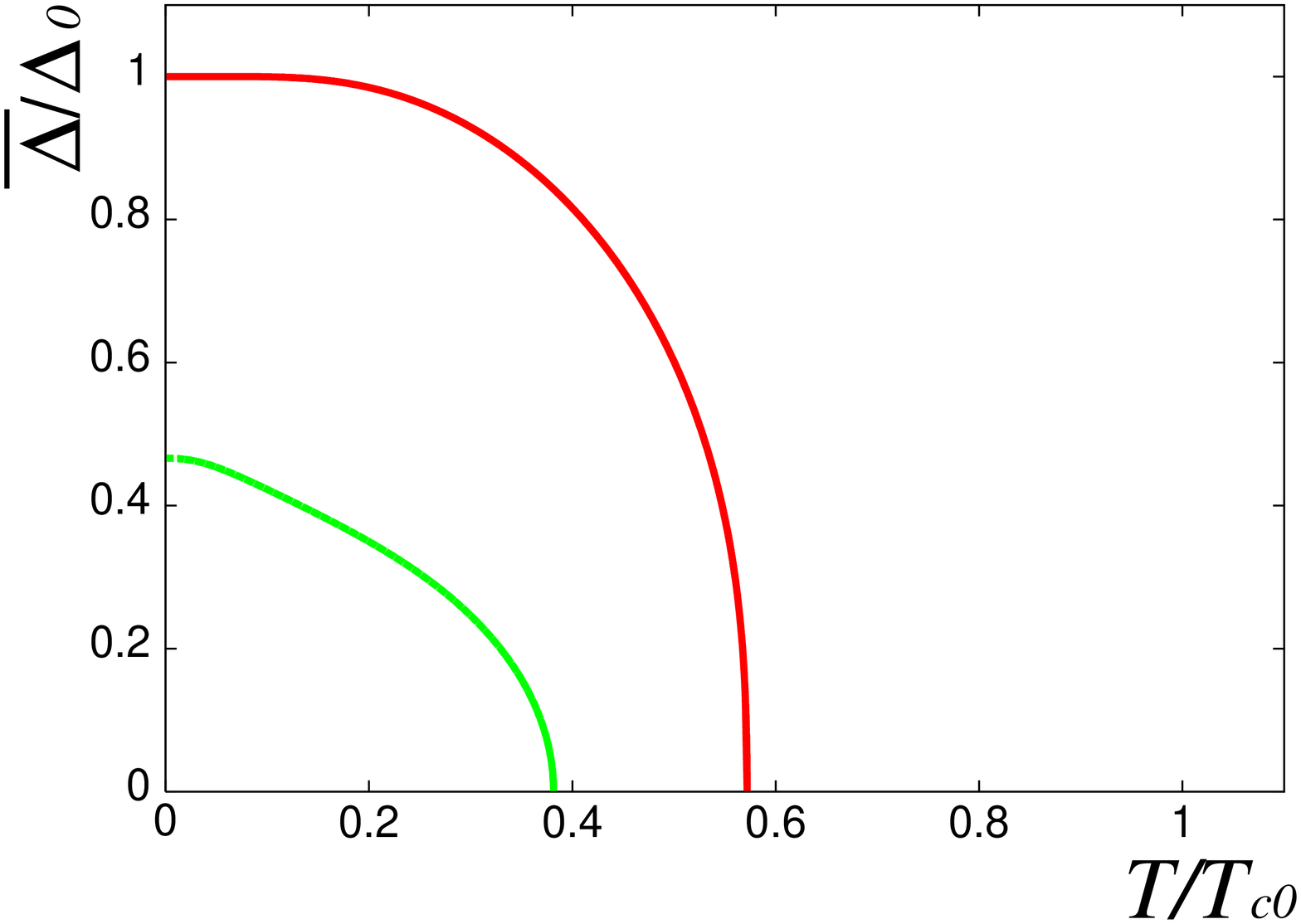}}\vspace{-5mm}
	\caption{The order parameter versus temperature for $V=1.213V_{10}$. The
 	solid lines correspond to the stable phase and the dashed lines correspond 
	to the unstable phase.}
	\label{GT@0.110}
}
\hfill
\parbox{\halftext}{\rotatebox[origin=c]{-90}{\includegraphics[scale=0.26]
	{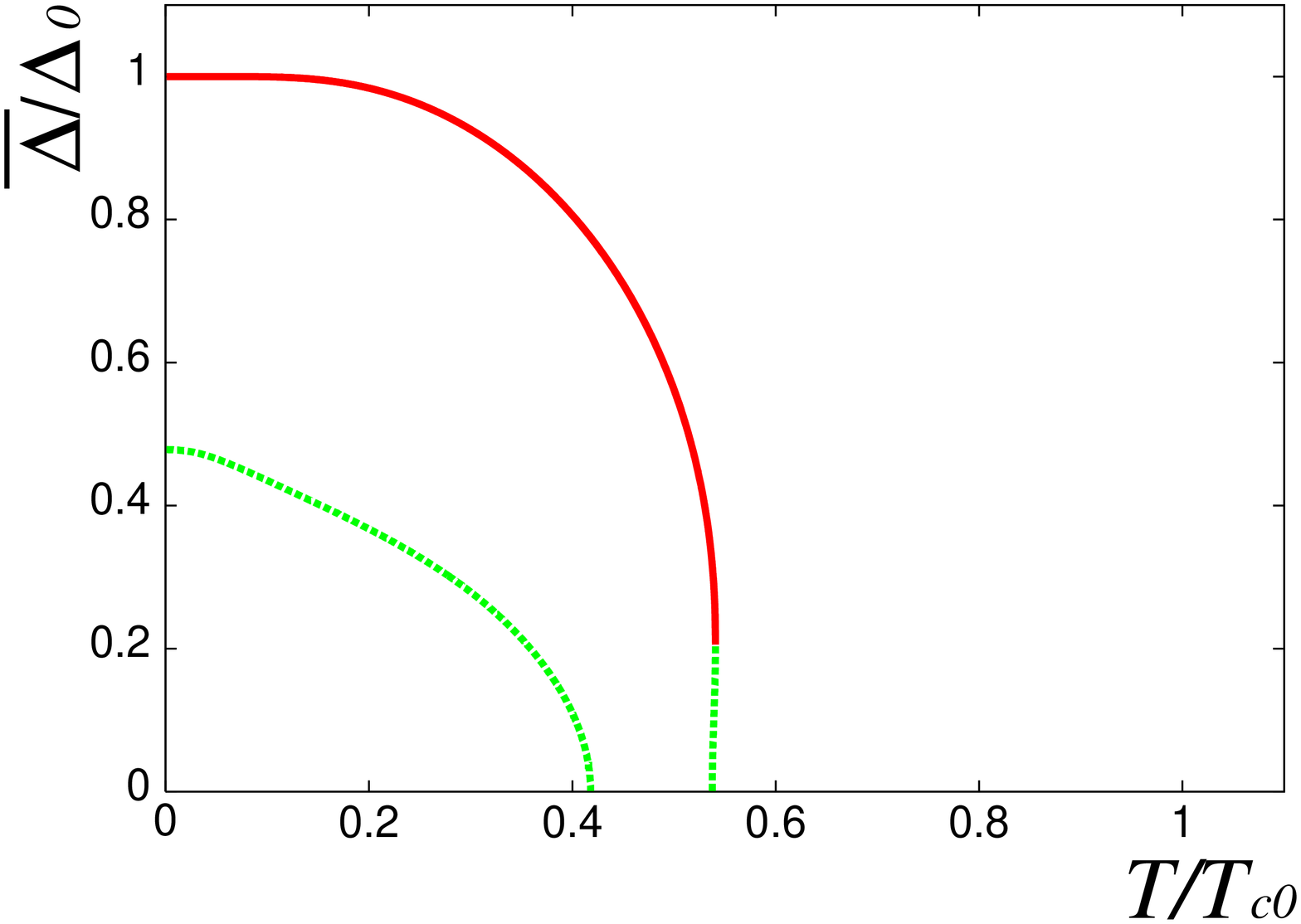}}\vspace{-5mm}
	\caption{The order parameter versus temperature for $V=1.224V_{10}$. The
 	solid lines correspond to the stable phase and the dashed lines correspond 
	to the unstable phase.}
	\label{GT@0.111}}
\end{figure}
\begin{figure}[h]
\parbox{\halftext}{\rotatebox[origin=c]{-90}{\includegraphics[scale=0.26]
	{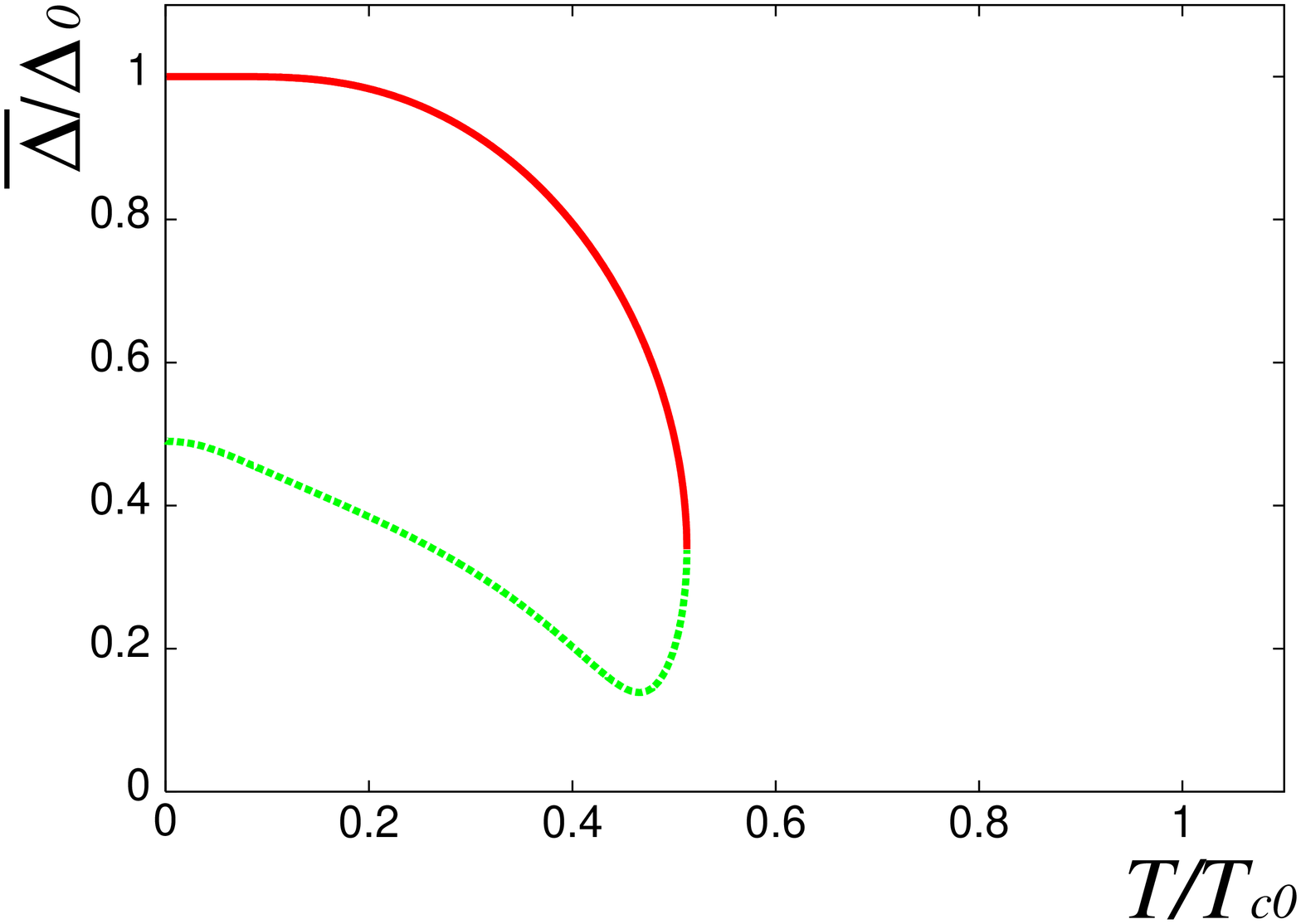}}\vspace{-5mm}
	\caption{The order parameter as a function of temperature for $V=1.235V_{10}$.
        The solid lines and dashed lines in this figure have 
        the same meaning as those in Figs.~\ref{vg@low} and Fig.~\ref{vi@low}.}
	\label{GT@0.112}
}
\hfill
\parbox{\halftext}{\rotatebox[origin=c]{-90}{\includegraphics[scale=0.26]
	{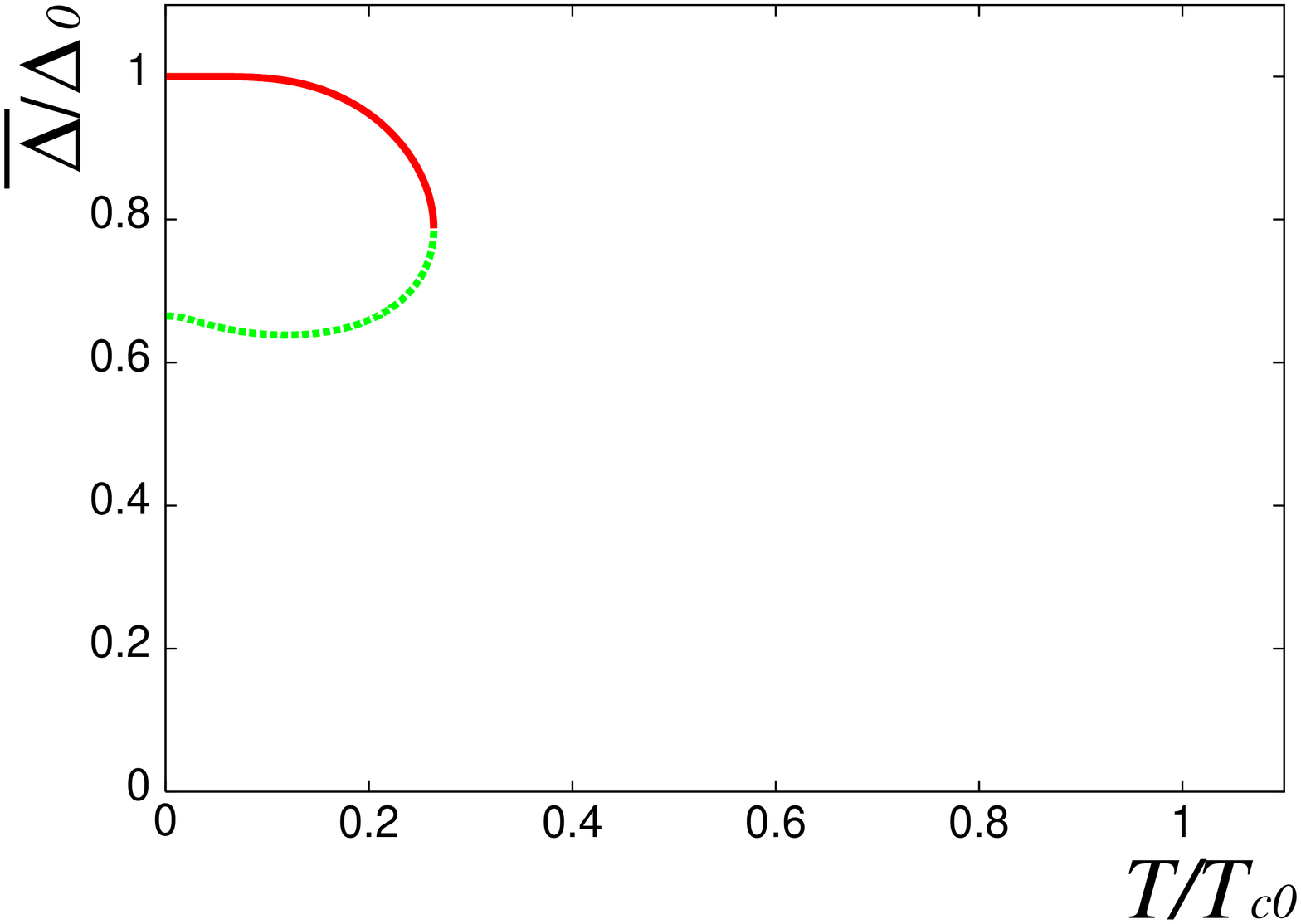}}\vspace{-5mm}
	\caption{The order parameter  as a function of temperature for $V=1.433V_{10}$.
        The solid lines and dashed lines in this figure have 
        the same meaning as those in Figs.~\ref{vg@low} and Fig.~\ref{vi@low}.}
	\label{GT@0.130}}
\end{figure}

\noindent
\underline{(B) Order Parameter}

Both the bias-voltage dependence of the order parameter and the current-voltage characteristics change continuously 
from those at zero temperature $T=0$ as shown in Figs.~\ref{vg@low}-\ref{vi@high} for $1/\lambda=4.8$.
When temperature is less than $T^*$ corresponding to the point P of Fig.~5, 
a voltage range $V_1(T)<|V|<V_2(T)$ exists where the order 
parameter is a triple-valued function of the bias voltage
$|V|$ (see Fig.~\ref{vg@low}) with 
an unstable middle branch 
and, hence, where the first-order phase transition is possible. 
The corresponding current-voltage characteristics are S-shaped, as shown in Fig.~\ref{vi@low}. 
Note that the small current observed at low bias voltage is due to thermally activated carriers.
On the other hand,
when the temperature is higher than $T^*$, the unstable branch disappears and the order parameter becomes a single-valued
function of the bias voltage (see Fig.~\ref{vg@high}). 
In this case, the current is a monotonically increasing
function of the bias voltage (see Fig.~\ref{vi@high}). 

The temperature dependence of the order parameter at constant bias voltage is shown in 
Figs.~\ref{GT@0.050}-\ref{GT@0.130}. 
At low bias voltage, the temperature dependence of the order parameter is similar to that in the absence of the bias 
(see Fig.~\ref{GT@0.050}). 
As the bias voltage increases, a lower-temperature branch corresponding to the unstable phase appears 
(see Fig.~\ref{GT@0.095}). 
In the temperature range where the unstable phase appears, the normal phase is stable in the sense of $\chi_N>0$, 
and the first-order 
phase transition between the stable ordered and normal phases is possible.
As the bias voltage increases 
further, the unstable 
branch approaches the stable branch and the two branches join (Figs.~\ref{GT@0.110}-\ref{GT@0.112}). 
Note that an unstable portion appears in the 
outer curve of Fig.~\ref{GT@0.111}. This is because the temperature $T^*$ is higher than 
the temperature where ${dV_1(T)\over dT}=0$.
For higher bias voltage, the stable ordered phase always co-exists with the normal phase, and the region 
where they co-exist shrinks with increasing
bias voltage (see Figs.~\ref{GT@0.112} and \ref{GT@0.130}). We remark that both the phase diagram (Fig.~\ref{PhaseDiagram}) 
and the temperature
dependence of the order parameter (Figs.~\ref{GT@0.112} and \ref{GT@0.130}) are similar to those for the nonequilibrium 
superconducting phase induced by
excess quasiparticles, which was studied by Scalapino {\it et al.}\cite{Scalapino1,Scalapino2}. 
This will be discussed in more detail in the last section.
\begin{wrapfigure}{r}{7cm}
\rotatebox[origin=c]{-90}{
\includegraphics[width=5cm]{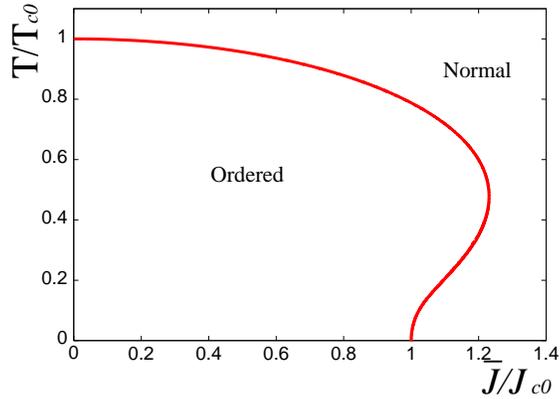}
}
\caption{Phase diagram at constant current. Only the second-order phase transition occurs at constant current.}
\label{PhaseDiagram2}
\end{wrapfigure}

\subsection{Phases at finite temperature at constant current \label{subsec. Finite2}}

\noindent
\underline{(A) Phase Diagram}

As in the zero-temperature case, at constant current, all the nontrivial solutions are stable in 
the sense of $\chi_I>0$, and the properties are drastically changed from those at constant bias voltage. 
First, only the ordered and normal phases exist and the phase transition between the two is always
of
second order. The corresponding boundary curve is given by (\ref{scaling2}):
\begin{equation}
\log\Big({\pi T_{c0}\over 2e^\gamma T}\Big)=
\phi\left({\pi T_{c0}\overline{J} \over 2e^\gamma J_{c0}T}\right)
\ ,
\label{scaling3}
\end{equation}
where $J_{c0}$ is the zero-temperature threshold current defined just above (\ref{GapCurrentZero}) (Fig.~\ref{PhaseDiagram2}).
From (\ref{scaling3}), the critical temperature $T_c(\overline{J})$ for small current and the threshold current $J_c(T)$ at 
low temperature are found to be
\begin{eqnarray}
T\equiv T_c(\overline{J})\simeq
T_{c0}\left\{1-{7\zeta(3)\over 16e^{2\gamma}}\left({\overline{J}\over J_{c0}}\right)^2 \right\}
\label{TcDec2}
\end{eqnarray}
and 
\begin{eqnarray}
|\overline{J}|\equiv J_c(T)
\simeq J_{c0}\Big\{1+{2e^{2\gamma}\over 3}\Big({T\over T_{c0}}\Big)^2\Big\}
\ .
\label{temp.scaling2}
\end{eqnarray}

The difference between the phase diagram at constant bias voltage and that at constant current
can be understood as follows. Since a larger order parameter implies a smaller current, the nontrivial
phase with larger order parameter in region B of Fig.~\ref{PhaseDiagram} corresponds to the phase 
with smaller current. 
As a result, the phase diagram in Fig.~\ref{PhaseDiagram2} does not have a region where more than one
phase is stable.
\bigskip

\begin{figure}[t]
\parbox{\halftext}{\rotatebox[origin=c]{-90}{\includegraphics[scale=0.26]
	{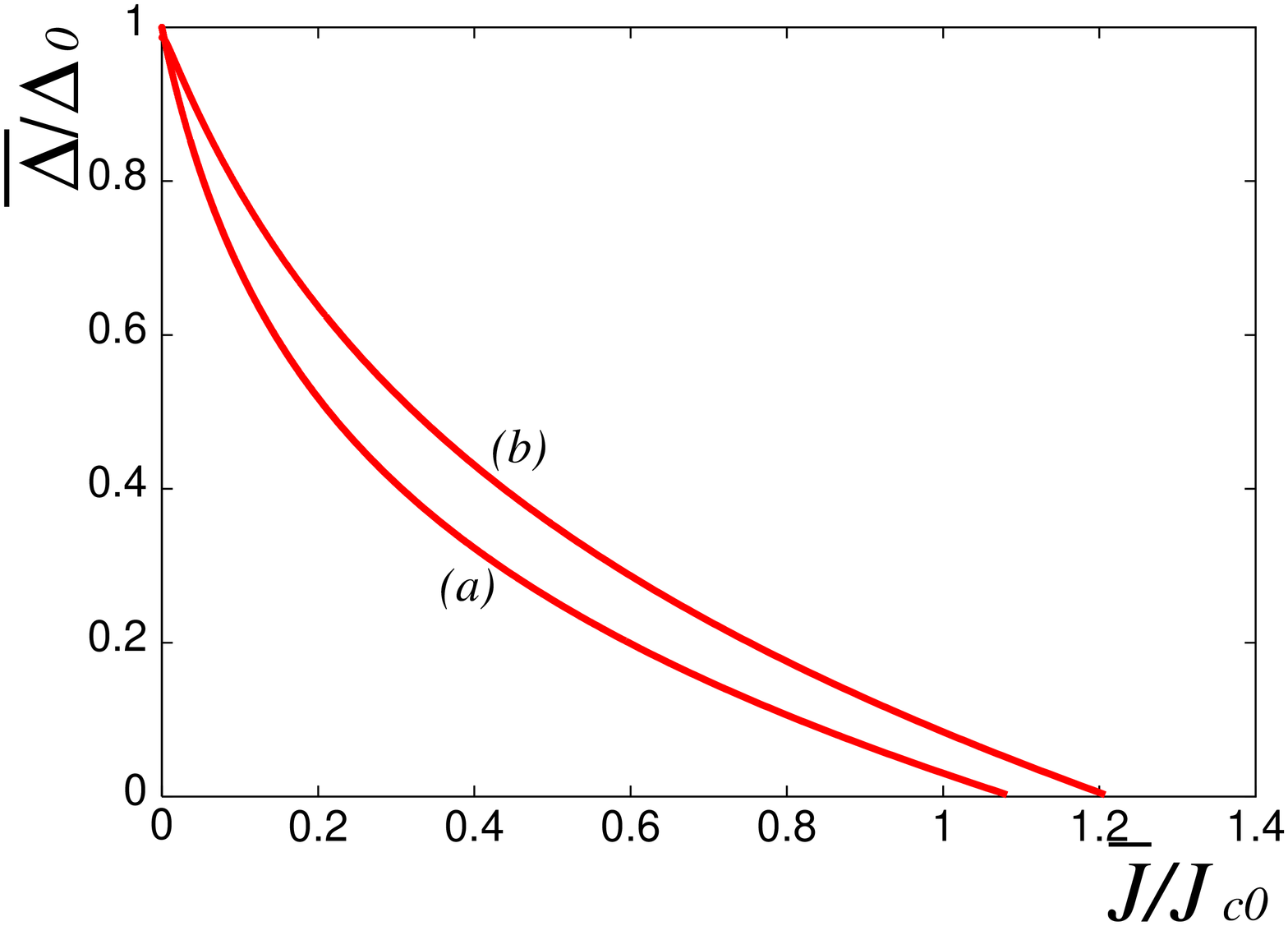}}
	\caption{The order parameter versus current for (a) $T=0.01\exp(1/2\lambda)|eV_{10}|$ $(=0.194 T_{c0})$
	and (b) $T=0.02\exp(1/2\lambda)|eV_{10}|$ $(=0.389 T_{c0})$. All phases shown in this figure are stable 
	 under constant current.}
	\label{ig@low}}
\hfill
\parbox{\halftext}{\rotatebox[origin=c]{-90}{\includegraphics[scale=0.26]
	{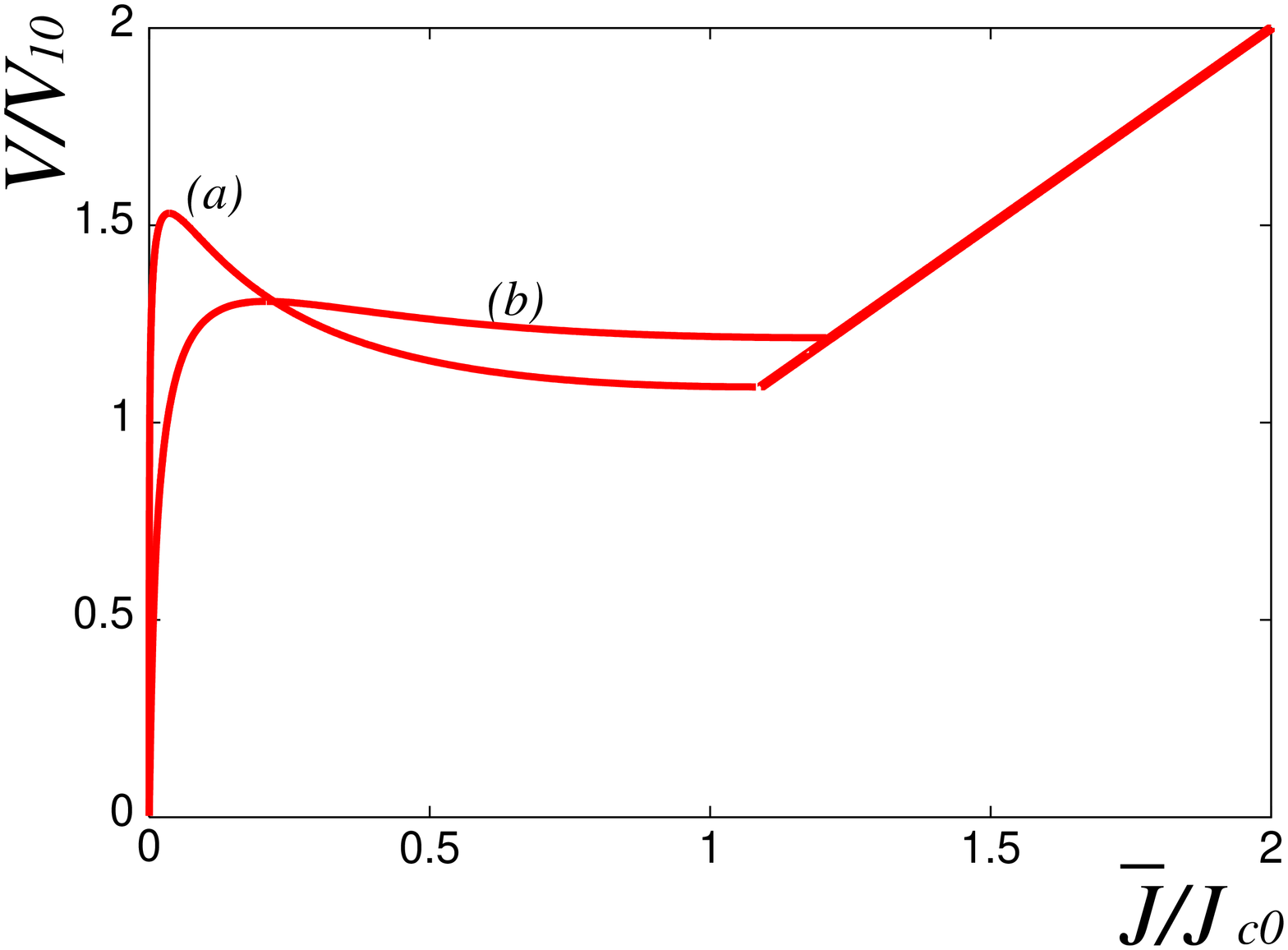}}
	\caption{Current versus bias voltage for (a) $T=0.01\exp(1/2\lambda)|eV_{10}|$ $(=0.194 T_{c0})$
	and (b) $T=0.02\exp(1/2\lambda)|eV_{10}|$ $(=0.389 T_{c0})$. All phases shown in this figure are stable 
	 under constant current.}
	\label{iv@low}}
\end{figure}
\begin{figure}[t]
\parbox{\halftext}{\rotatebox[origin=c]{-90}{\includegraphics[scale=0.26]
	{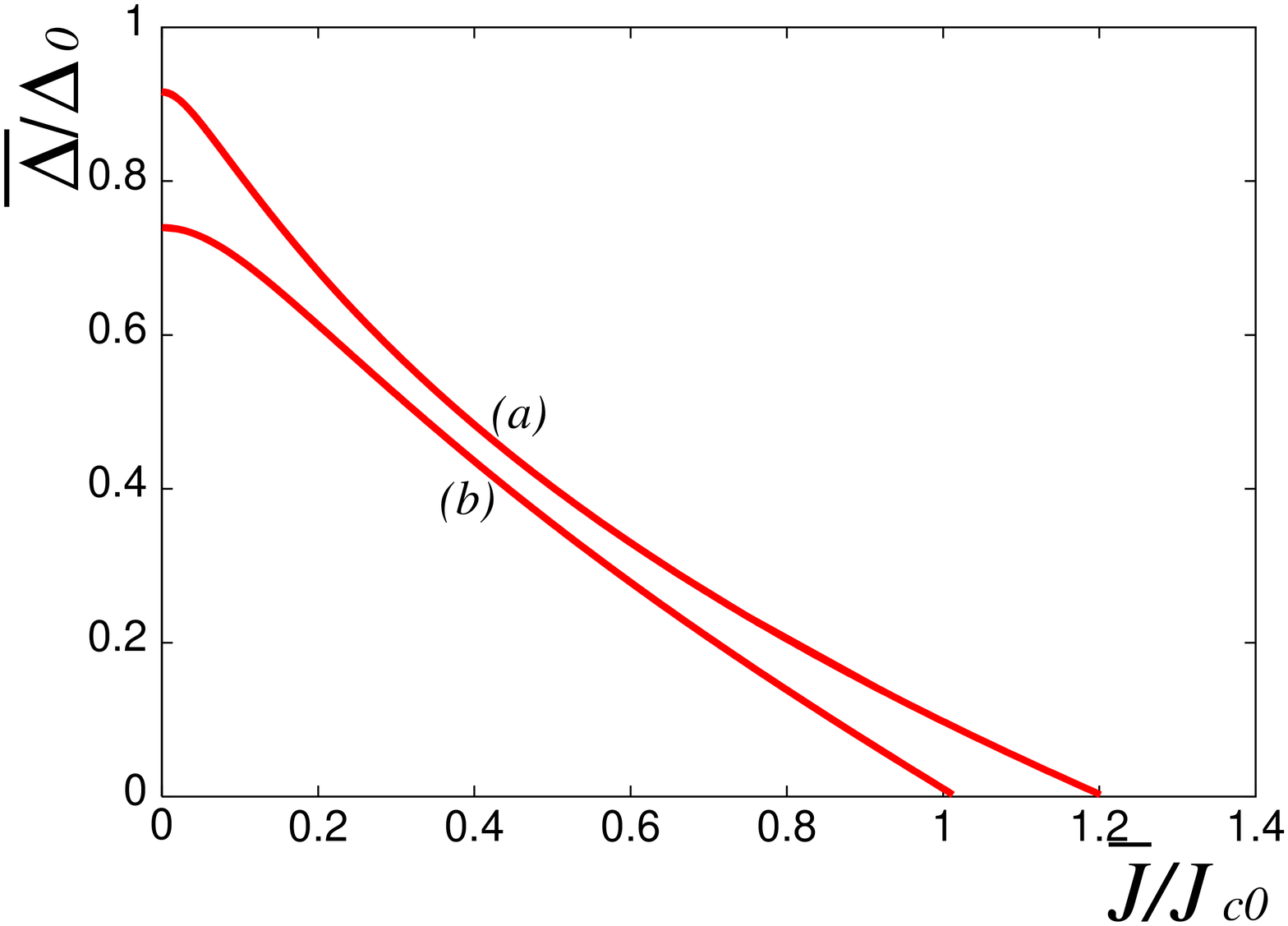}}
	\caption{The order parameter versus current for (a) $T=0.03\exp(1/2\lambda)|eV_{10}|$ $(=0.583 T_{c0})$
	and (b) $T=0.04\exp(1/2\lambda)|eV_{10}|$ $(=0.778 T_{c0})$. All phases shown in this figure are stable 
	 under constant current.}
	\label{ig@high}}
\hfill
\parbox{\halftext}{\rotatebox[origin=c]{-90}{\includegraphics[scale=0.26]
	{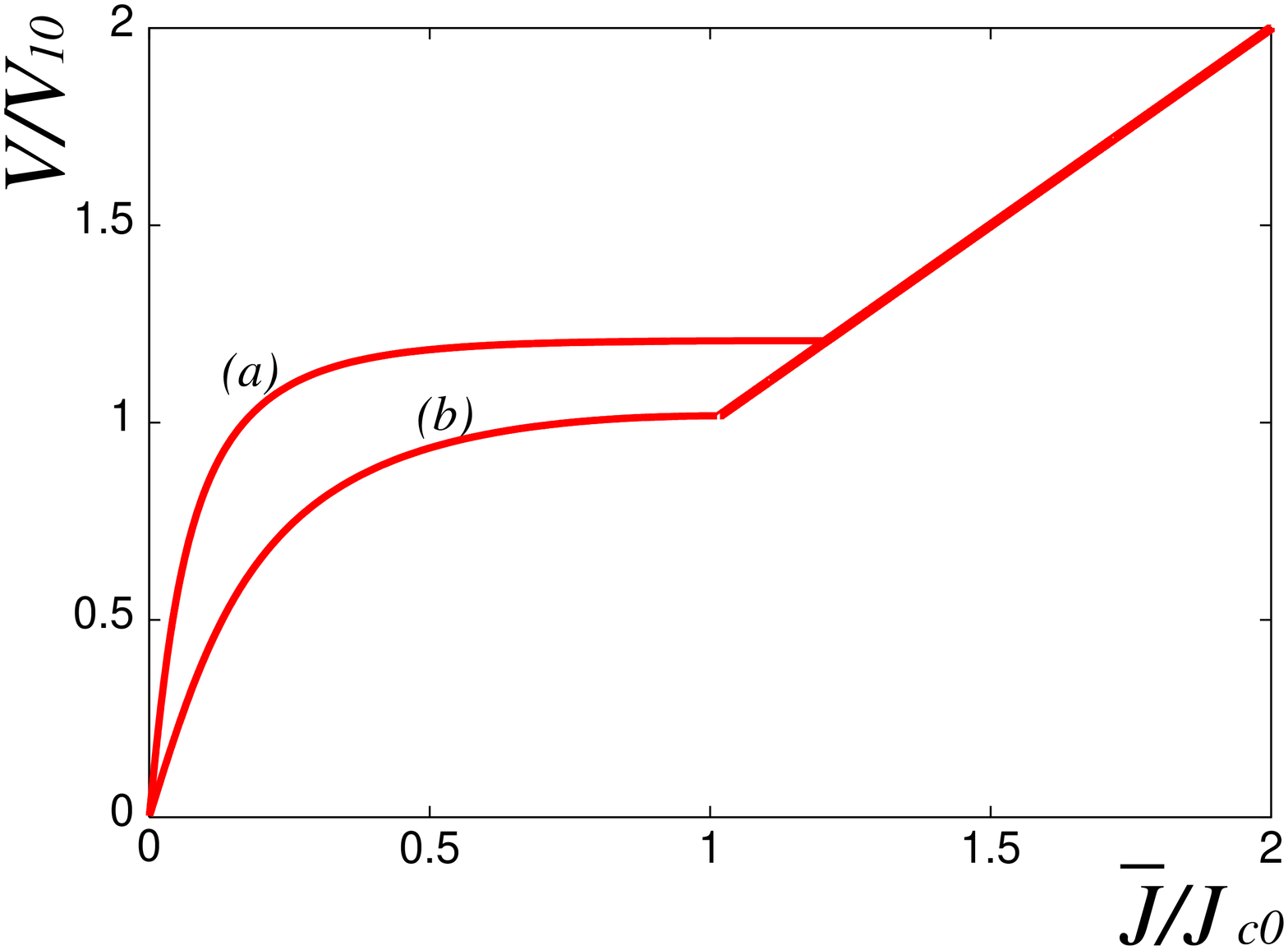}}
	\caption{Current versus bias voltage for (a) $T=0.03\exp(1/2\lambda)|eV_{10}|$ $(=0.583 T_{c0})$
	and (b) $T=0.04\exp(1/2\lambda)|eV_{10}|$ $(=0.778 T_{c0})$. All phases shown in this figure are stable 
	 under constant current.}
	\label{iv@high}}
\end{figure}

\noindent
\underline{(B) Order Parameter}

As before, when temperature increases, both the current dependence of the order parameter and the voltage-current 
characteristics change continuously from those at zero temperature. 
At any temperature, the order parameter is a monotonically decreasing function of the current (cf.
Figs.~\ref{ig@low} and \ref{ig@high}). 
For $|\overline{J}|\sim 0$, the decrease in the order parameter is found to be proportional to the squared current, but the quadratic region is not 
visible at lower temperature $T\le 0.6\times T_{c0}\simeq T^*$, where the 
decrease in the
order parameter is approximately proportional
to the current. 
The corresponding voltage-current characteristics behave as shown in 
Figs.~\ref{iv@low} and \ref{iv@high}. 
We remark that negative differential conductivity appears only at temperatures lower than $T^*$,
and otherwise the differential conductivity is positive.
\begin{wrapfigure}{r}{7cm}
\rotatebox[origin=c]{-90}{
\includegraphics[width=5cm]{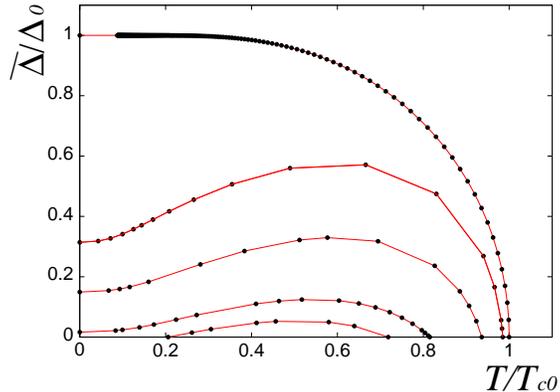}
}
\caption{The order parameter as a function of temperature for $\overline{J}/J_{c0}=0$, 0.3, 0.6, 0.95, and 1.10 
from top to bottom. 
To draw this figure, we use an approximate equation of (\ref{SelfUniNum}) similar to (\ref{scaling}).
For computational reasons, the numbers of data are limited for $\overline{J}/J_{c0}=$0.3, 0.6, 0.95 and 1.10.
}
\label{GT@I}
\end{wrapfigure}

The temperature dependence of the order parameter at constant current 
is shown in Fig.~\ref{GT@I}. 
The order parameter is reduced by the presence of the current. 
Even for a smaller current  (see the cases $\overline{J}/J_{c0}=0.3$ and 0.6 in Fig.~\ref{GT@I}), 
the lower-temperature part is more suppressed than the higher-temperature part.
At the threshold current $\overline{J}=J_{c0}$, the order parameter at zero temperature vanishes
(see the case $\overline{J}/J_{c0}=0.95$ in Fig.~\ref{GT@I}).
When the current exceeds the threshold, re-entrance to the normal phase appears at low temperature.
As the current increases, the 
temperature range with 
nonvanishing order parameter shrinks and eventually
vanishes (see the case $\overline{J}/J_{c0}=1.10$ in Fig.~\ref{GT@I}).

\section{Summary and Discussions}
We have studied the nonequilibrium Peierls transition in a TLM chain connected to two reservoirs at different 
chemical potentials (their difference corresponds to the bias voltage) by combining a mean-field approximation and 
formula 
(\ref{ChNESS}), which characterizes a nonequilibrium steady state and
is an outcome of the algebraic field-theoretical 
approach to nonequilibrium statistical mechanics. The averaged 
lattice distortion serves as an order parameter, and its self-consistent equation is obtained by averaging the equation of motion 
of the lattice distortion with respect to a nonequilibrium steady state. 
When the bias voltage and temperature are chosen as control parameters, three parameter regions are distinguished: 
region A where a single stable ordered phase is possible,
region B where stable normal, stable ordered, and unstable ordered phases are possible 
and region C where only the stable normal phase is possible (cf. Fig.~5). 
The transition between regions A
and C is of
second order, and the transition temperature decreases with increasing bias voltage.
A first-order phase transition between normal and ordered phases may occur in region B. 
In regions A and C, the current is a single-valued function of the bias voltage, and in region B, the current-voltage
characteristics are S-shaped (i.e. negative differential conductivity exists).
In contrast, when the current and temperature are chosen as control parameters, all the nontrivial solutions of 
the self-consistent equation
become stable in the sense 
of linear stability.
The phase transition between the ordered
and normal phases is always 
of second order, and re-entrant behaviour is seen for currents larger than the threshold value, $J_{c0}$.
Negative differential conductivity appears only when the temperature is lower than a certain value $T^*$.
We remark that, as in the equilibrium case\cite{Gruner}, the mean 
field approach is expected to provide a qualitatively correct description of the nonequilibrium 
phase transition in quasi-1D systems, although the 1D order may be destroyed by fluctuations.  

As mentioned in the previous section, 
the phase diagram, particularly the possibility of the re-entrant behaviour, and the
temperature dependence of the order parameter at higher bias voltage
are similar to those of the nonequilibrium superconducting
phase induced by excess quasiparticles, which were studied by Scalapino {\it et al.}\cite{Scalapino1,Scalapino2}.
This can be understood because of the similarity between the self-consistent equations in the two cases.
Indeed, let $E$ be $\sqrt{\epsilon^2+{\overline\Delta}^2}$, then the self-consistent equation (\ref{SelfUniNum}) 
is rewritten as 
\begin{eqnarray}
{1\over \lambda}=\int_{-\hbar \omega_c}^{\hbar\omega_c} {d\epsilon\over E}\Big\{\tanh{1\over 2T}\Big(E-{eV\over 2}\Big)
+\tanh{1\over 2T}\Big(E+{eV\over 2}\Big)\Big\}
\ , 
\end{eqnarray}
which reduces to the self-consistent equation of Scalapino {\it et al.} [cf. Eq.(6) of Ref.~\citen{Scalapino1}] if 
the second term 
is dropped and $eV/2$ is replaced with the effective chemical potential $\mu^*$.
In view of this similarity, one can interpret the suppression of the charge-density-wave order 
induced by the bias voltage 
(equivalently by the current) as being due to excess electrons coming from the two reservoirs.
However, since the two self-consistent equations are not exactly the same, the two systems are different in their
properties at lower bias voltage (namely, in the temperature dependence of the order parameter and 
the possibility of the second-order 
phase transition).

As is well known, systems with density waves may exhibit nonlinear conduction 
due to density-wave sliding\cite{Gruner}. On the other hand, because of the above observation and 
the similarity between the 
mean-field approximation for superconductors and that for density 
waves, the current-induced suppression of order discussed here 
is generally expected for systems with density waves. However, except for cases where density waves are strongly pinned, 
the current-induced suppression of order may not be observed 
because it is related to the amplitude degrees of freedom, while
the sliding is related to the easily excitable phase degrees of freedom. 

We remark that the mean-field approximation of the TLM chain is equivalent to the mean-field approximation of the half-filled
charge order in the spinless extended Hubbard chain, the Hamiltonian of which is given by
\begin{eqnarray}\label{ExtHubbard}
H_{\rm CO}=-t_0\sum_{j=0}^{L-1}\{C_{j+1}^\dag C_j+C_j^\dag C_{j+1}\}+U\sum_{j=0}^{L-1} n_{j+1}n_j
-U\sum_{j=0}^Ln_j \ ,
\end{eqnarray}
where $n_j=C_j^\dag C_j$ is the number operator 
of the spinless fermions
at site $j$.
Indeed, in the continuum limit discussed in Appendix~A [cf. (\ref{ContinuousLimit})], 
the mean-field equations become (\ref{NESSMF1}) and (\ref{Self1}) with
replacement $\sigma_x\to \sigma_z$ and $\lambda\to U/(\pi t_0)$, where the order parameter is proportional to the charge 
disproportion between even and odd sites. 
Hence, as in the open TLM chain,
the charge order in the open extended Hubbard chain is suppressed by current. 
This observation suggests that current-induced suppression 
could be 
a possible origin of 
the nonlinear conduction, which is different from phenomena such as sliding density waves\cite{Gruner}, strong impurity 
scattering in a
Tomonaga-Luttinger liquid\cite{Egger}, dielectric breakdown of Mott insulators\cite{Oka} or 
the Kosterilitz-Thouless transition\cite{Takahide}.

It is then interesting to compare the present results with the experiments by Terasaki and his 
co-workers\cite{Inagaki,Sawano1,Wata,Sawano2} on 
the charge order in the organic 
conductors 
$\theta$-(BEDT-TTF)$_2 {\rm C}{\rm s}M$(SCN)$_4$
($M=$Zn, Co, Co$_{0.7}$Zn$_{0.3}$).
The existence of hysteresis and spontaneous oscillation at constant bias voltage and negative differential conductivity at
constant current reported in Ref.~\citen{Sawano1} is consistent with the present results. 
In Ref.~\citen{Sawano2}, it was shown that the current-induced decrease in the order parameter 
is proportional to the current and, 
this observation agrees with the current dependence of the order parameter at temperatures where nonlinear conduction is possible
(cf. Figs.~\ref{ig@low} and \ref{ig@high}). 
Also, according to the present analysis, negative differential conductivity appears only when $\lambda\sim U/(\pi t_0)$
is not too large. This seems to imply that systems with weaker charge ordering are more favourable for 
negative differential conductivity,
and further imply that, because of the fragility of their charge order, the organic 
conductors
mentioned above would be such systems.
Moreover, in these materials, the charge order has no long-range order in the phase mode, and thus 
excitations in the amplitude mode instead of
charge-order sliding would be responsible for nonlinear conductivity.
These observations suggest that current-induced suppression of the charge order is the origin 
of the phenomena investigated
by Terasaki and co-workers.\cite{Inagaki,Sawano1,Wata,Sawano2} 
More quantitative analysis based on a realistic model of the organic 
conductors
will be studied elsewhere.

Moreover,
since the extended Hubbard model (\ref{ExtHubbard}) is a typical model of strongly
correlated systems, the present analysis would provide some insight into the
negative differential conductivity recently reported 
in strongly correlated
systems\cite{Egger,Oka,Casati,Casati2}. In particular, as the extended Hubbard chain
(\ref{ExtHubbard}) is equivalent to the XXZ chain via the Jordan-Wigner transformation, 
the negative differential conductivity in the nonequilibrium XXZ model found by Benenti 
{\it et al.}\cite{Casati} should be approximately understood in terms of the present results. 
However, further investigation is necessary because in the work of Benenti {\it et al}., 
the system is driven to a nonequilibrium steady state by 
stochastic activation of the boundary spins, not
through coupling with infinitely extended reservoirs.

\section*{Acknowledgements}
The authors thank T. Prosen, G. Benenti, G. Casati,
Y. Matsunaga, Baowen Li, Bambi Hu, K. Nakamura, A. Sugita for fruitful discussions.
This work is partially supported by a Grant-in-Aid for Scientific
Research (Nos. 17340114, 16076213, and 17540365) from the Japan Society of 
the Promotion of Science,
for the ``Academic Frontier'' Project at Waseda University 
and the 21st Century COE Program at Waseda University ``Holistic
Research and Education Center for Physics of Self-organization Systems''
both from the Ministry of Education, Culture, Sports, Science and
Technology of Japan.

\appendix
\section{Derivation of a continuous model \label{ap.SSH}}

In this appendix, we derive the continuous open TLM model from the discrete open SSH model.
Hamiltonian of our system is consist of the SSH part ($H_S$), two reservoirs ($H_B$), 
and the interaction between SSH part and reservoirs ($V$): 
\begin{eqnarray}
H&=&H_{S} + V+  H_B
\nonumber
\\
H_S&=&-\sum_{\sigma}\sum_{n=-1}^{L+1}
\left( t_{n+1,n} C_{n+1 \sigma}^\dag C_{n \sigma} + ({\rm{h.c.}}) \right)+
\frac{K}{2}\sum_{n=-1}^{L} (y_{n+1}-y_n)^2+
\frac{M}{2}\sum_{n=0}^{L} \dot{y}_n^2
\nonumber
\\
V&=&\sum_{\sigma}\integ{k}
\hbar  \left(\bar{v}_k
C_{0 \sigma}^{\dag}a_{k \sigma} + 
\bar{w}_k
C_{L \sigma}^{\dag}b_{k \sigma} + ({\rm{h.c.}}) 
\right)	,\ 
\nonumber
\\
H_B&=&\sum_{\sigma}\integ{k}
( \hbar \omega_{kL} a_{k \sigma}^{\dag}a_{k \sigma} + 
\hbar \omega_{kR} b_{k \sigma}^\dag b_{k \sigma} ) 
\ ,
\label{DHamiltonian}
\end{eqnarray}
where $C_{n \sigma}$ denotes the annihilation operator of an electron at the $n$th site with spin $\sigma$ 
(cf. $C_{-1 \sigma}\equiv 0,\ C_{L+1 \sigma}\equiv 0$), 
$a_{k\sigma} (b_{k\sigma})$ denotes the annihilation operator of the left (right) reservoir with wave number $k$ and 
spin $\sigma$, and $y_n$ denotes a 
lattice displacement of the $n$th site, respectively.
Su, Schrieffer, and Heeger~\cite{SSH} assumed that $t_{n+1,n}$ is a linear function of the lattice displacement:
\begin{eqnarray*}
t_{n+1, n}\equiv t_0-\alpha(y_{n+1}-y_n)
\ .
\end{eqnarray*}
Takayama {\it et al.}\cite{TLM} approximated the dispersion relation of 
electrons as $-2t_0
\cos[(k\pm k_F)a]\approx\pm v_f k$, and introduced the left/right moving
electron fields $\psi_{L \sigma}(2na)$/$\psi_{R \sigma}(2na)$, and 
$\Delta(na)=(-1)^{n}4\alpha y_n$
with $a$ the lattice constant. 
Then, by assuming $L\equiv -1$ (mod 4) and considering $a$ to be very small, the discrete Hamiltonian~(\ref{DHamiltonian}) reads as
\begin{eqnarray*}
H^{\rm (e)}_S &\simeq&
\sum_{\sigma} \int_0^l dx
\left( \psi_{L \sigma}^\dag(x) , \psi_{R \sigma}^\dag(x) \right)
\left[-2iat_0
\sigma_y\frac{\partial }{\partial x}+ 
\sigma_x \Delta (x)
\right]
\binom{\psi_{L \sigma}(x)}{\psi_{R \sigma}(x)}
\nonumber
\\
H^{\rm (ph)}_S &\simeq&
\frac{K}{8\alpha^2 a}\int_0^l dx \Delta(x)^2+
\int_0^l dx \frac{M}{32\alpha^2 a}\dot{\Delta}(x)^2
\nonumber
\\
V&\simeq& \sqrt{a} \sum_\sigma 
\int dk\  \hbar  \Big[\bar{v}_k
\left(-i\psi_{L \sigma}^\dag (a)+\psi_{R \sigma}^\dag (a)\right)
a_{k \sigma} + \bar{w}_k
\left(\psi_{L \sigma}^\dag (l)-i\psi_{R \sigma}^\dag (l)\right)
b_{k \sigma}
\Big] + (h.c.)
\nonumber
\\
H_B&=&
\sum_\sigma \integ{k}
( \hbar\omega_{kL} a_{k \sigma}^{\dag}a_{k \sigma} + 
\hbar\omega_{kR} b_{k \sigma}^\dag b_{k \sigma} )
\ .
\end{eqnarray*}
Moreover, $C_{-1 \sigma}\equiv 0$ and $C_{L+1 \sigma}\equiv 0$ lead to
\begin{eqnarray*}
\psi_{L \sigma}(0)+i\psi_{R \sigma}(0)=0,\ i\psi_{L \sigma}(l)+\psi_{R \sigma}(l)=0
\ ,
\label{boundary}
\end{eqnarray*}
where $\ell$ is defined by $\ell=a+La$. 
We remark that case of $L\equiv +1$ (mod 4) leads to essentially the same results.
To make the boundary condition simpler, we introduce $d_\sigma(x)$ and $e_\sigma(x)$ by
\begin{eqnarray}\label{ContinuousLimit}
C_{2n-1 \sigma}&=&
(-1)^n \sqrt{2a}\,d_\sigma (2na)=(-1)^n \sqrt{a}\big[\psi_{L \sigma} (2na)+i\psi_{R \sigma} (2na)\big]
\nonumber
\\
C_{2n \sigma}&=&
(-1)^n \sqrt{2a} e_\sigma (2na)=(-1)^n \sqrt{a}\big[i\psi_{L \sigma} (2na)
+\psi_{R \sigma}(2na)\big]
\ .
\end{eqnarray}
By using these fields, 
the momentum $\Pi (x)\equiv\dot{\Delta}(x)$ conjugate to ${\Delta}(x)$,
the Fermi velocity $v \equiv 2at_0/\hbar$, dimensionless coupling constant $\lambda\equiv 4\alpha^2 a/\pi\hbar vK$, 
phonon frequency $\omega_0\equiv\sqrt{4K/M}$, and the matrix elements $v_k\equiv \sqrt{2a}\bar{v}_k$
and $w_k\equiv \sqrt{2a}\bar{w}_k$, 
the Hamiltonians (\ref{HamHS}), (\ref{HamB}), and (\ref{HamInt}), and the boundary condition~(\ref{TLMboundary}) in \S2 are obtained.

\section{Normal modes \label{ap.normal}}

In this appendix, we derive normal modes of the mean-field Hamiltonian $H_{\rm\small MF}$.
Let $\{\phi_\lambda (x)\}_\lambda$ be a 
complete orthonormal solution of the 
eigenvalue problem:
\begin{eqnarray}
&&\left[-i \hbar v
\sigma_y\frac{\partial }{\partial x}+ 
\Delta (x) \sigma_x
\right]
\phi_\lambda(x)=\hbar\epsilon_\lambda\phi_\lambda(x)\ ,
\label{eigen1}
\\
&&\phi_\lambda (x)\equiv\binom{\phi_\lambda^+(x)}{\phi_\lambda^-(x)},\ \ 
\phi_\lambda^+(0)=0,\ \ \phi_\lambda^-(\ell)=0
\ ,
\label{eigen2}
\end{eqnarray}
and we expand the electron field
\begin{eqnarray}
&&\Psi_\sigma(x) = \binom{d_\sigma(x)}{e_\sigma(x)}=
\sum_{\lambda, \sigma}\phi_\lambda(x) f_{\lambda \sigma},\ \ 
\label{eigenExp}
\end{eqnarray}
where $\{f_{\lambda \sigma}, f_{\lambda' \sigma'}^\dag\}=
\delta_{\lambda,\lambda'} \delta_{\sigma,\sigma'}$.
In terms of $f_{\lambda\sigma}$, the mean-field Hamiltonian reads
\begin{eqnarray*}
{1\over \hbar}H_{\rm\small MF}&=&\sum_{\lambda,\sigma}\epsilon_\lambda 
f_{\lambda\sigma}^\dag f_{\lambda\sigma}
+\integ{{\bf k}} (\omega_{kL} a_{{\bf k}\sigma}^\dag a_{{\bf k}\sigma}+\omega_{kR} b_{{\bf k}\sigma}^\dag b_{{\bf k}\sigma})
\nonumber \\
&&+\sum_{\lambda ,\sigma}
\integ{{\bf k}}\ \left\{
\left( \phi_\lambda^-(0) v_{\bf k}^*  a^\dag_{{\bf k}\sigma} + \phi_\lambda^+(\ell) w_{\bf k}^* 
b^\dag_{{\bf k}\sigma} \right)f_{\lambda \sigma} + (h.c.)\right\} 
\ .
\end{eqnarray*}
Since $H_{\rm\small HF}$ is bilinear with respect to field operators, the incoming fields are
linear combinations of $a_{{\bf k}\sigma}, b_{{\bf k}\sigma}$ and $f_{\lambda\sigma}$:
\begin{eqnarray}
\alpha_{{\bf k}\sigma}&=&a_{{\bf k}\sigma} + \sum_\lambda h^{\bf k}_\lambda 
f_{\lambda\sigma}
+\integ{{\bf k}'} \left(
m^{\bf k}_{{\bf k}'}a_{{\bf k}'\sigma}+n^{\bf k}_{{\bf k}'}b_{{\bf k}'\sigma}
\right) \ ,
\label{in form}
\end{eqnarray}
By substituting it into $[\alpha_{{\bf k}\sigma},H_{\rm\small MF}]/\hbar=\omega_{kL}\alpha_{{\bf k}\sigma}$
and comparing term by term, one has
\begin{eqnarray}
\alpha_{{\bf k}\sigma}&=&a_{{\bf k}\sigma} + \sum_\lambda h^{\bf k}_\lambda 
f_{\lambda\sigma}
+\integ{{\bf k}'} \Big(
\frac{v_{{\bf k}'}A^-_{\bf k}(0) a_{{\bf k}'\sigma}}{\omega_{kL}-\omega_{k'L}\pm i0} 
+\frac{w_{{\bf k}'}A^+_{\bf k}(\ell) b_{{\bf k}'\sigma}}{\omega_{kL}-\omega_{k'R}\pm i0}
\Big) \ ,
\label{InAlpha}
\\
h^{\bf k}_\lambda
&=&\frac{\phi^-_\lambda(0)}{\omega_{kL}-\epsilon_\lambda} \big\{ v_{\bf k}^*
+A_{\bf k}^-(0) \xi_\pm(\omega_{kL})\big\}+
\frac{\phi^+_\lambda(\ell)}{\omega_{kL}-\epsilon_\lambda}A_{\bf k}^+(\ell)
\eta_\pm(\omega_{kL}) \ ,
\label{h-lambda1}
\end{eqnarray}
where $A^\rho_{\bf k}(x)=\sum_{\lambda}\phi^\rho_\lambda(x)^* h^{\bf k}_\lambda$
($\rho=\pm$; $x=0,\ell$) and 
\begin{eqnarray}
\xi_\pm(z)&\equiv& \integ{{\bf k}'}\frac{|v_{{\bf k}'}|^2}{z-\omega_{k'L}\pm i0}
,\ \ \ 
\eta_\pm(z)\equiv \integ{{\bf k}'}\frac{|w_{{\bf k}'}|^2}{z-\omega_{k'R}\pm i0}
\ .
\end{eqnarray}
Substituting (\ref{h-lambda1}) into the definition of $A^\rho_{\bf k}(x)$, one obtains
a linear equation for $A^-_{\bf k}(0)$ and $A^+_{\bf k}(\ell)$ and its solution is
\begin{eqnarray}
A_{\bf k}^+(\ell)&=&v_{\bf k}^* {g_{-+}(0,\ell:\omega_{kL})\over \Lambda_\pm(\omega_{kL})}
\ ,
\\
v_{\bf k}^*+\xi_\pm(\omega_{kL})A_{\bf k}^-(0)&=&
v_{\bf k}^* {1-\eta_{\pm}(\omega_{kL})g_{++}(\ell,\ell:\omega_{kL})\over \Lambda_\pm(\omega_{kL})}
\ ,
\end{eqnarray}
where 
\begin{eqnarray}
&&\Lambda_\pm(z)=1-\xi_\pm(z)g_{--}(0,0;z)-\eta_\pm(z)g_{++}(\ell,\ell;z)
\nonumber \\
&&~~~~+\xi_\pm(z)\eta_\pm(z)
\{g_{++}(\ell,\ell;z)g_{--}(0,0;z)-g_{+-}(\ell,0;z)g_{-+}(0,\ell;z)
\}
\end{eqnarray}
and $g_{\rho\rho'}(x,y:z)$ is the $\rho\rho'$-component of the Green function:
\begin{equation}
G(x,y;z)=
\left(\begin{matrix}
g_{++}(x,y;z) &g_{+-}(x,y;z)\cr
g_{-+}(x,y;z) &g_{--}(x,y;z)\cr
\end{matrix}\right)
\equiv \sum_\lambda {\phi_\lambda(x)\phi_\lambda(y)^\dag\over z-\epsilon_\lambda}
\ .
\end{equation}
As a result of the completeness of the eigenfunctions $\phi_\lambda(x)$, the Green function satisfies
\begin{eqnarray}
&&\left[-i \hbar v
\sigma_y\frac{\partial }{\partial x}+ 
\Delta (x) \sigma_x
\right]G(x,y:z)
=\hbar z G(x,y;z)-\hbar {\bf 1}\delta(x-y)
\ ,
\\
&&
g_{++}(0,y;z)=g_{+-}(0,y;z)=g_{-+}(\ell,y;z)=g_{--}(\ell,y;z)=0
\ ,
\end{eqnarray}
where $\bf 1$ stands for the 2$\times$2 unit matrix.
Similarly, we have
\begin{eqnarray}
\beta_{{\bf k}\sigma}&=&b_{{\bf k}\sigma} + \sum_\lambda {\widetilde h}^{\bf k}_\lambda 
f_{\lambda\sigma}
+\integ{{\bf k}'} \Big(
\frac{v_{{\bf k}'}B^-_{\bf k}(0) a_{{\bf k}'\sigma}}{\omega_{kR}-\omega_{k'L}\pm i0} 
+\frac{w_{{\bf k}'}B^+_{\bf k}(\ell) b_{{\bf k}'\sigma}}{\omega_{kR}-\omega_{k'R}\pm i0}
\Big) \ ,
\label{InBeta}
\\
{\widetilde h}^{\bf k}_\lambda
&=&\frac{\phi^-_\lambda(0)}{\omega_{kR}-\epsilon_\lambda} B_{\bf k}^-(0)
\xi_\pm(\omega_{kR}) +
\frac{\phi^+_\lambda(\ell)}{\omega_{kR}-\epsilon_\lambda}
\big\{ w_{\bf k}^*
+B_{\bf k}^+(\ell) \eta_\pm(\omega_{kR})\big\}
  \ ,
\label{h-lambda}
\end{eqnarray}
where
$B^\rho_{\bf k}(x)=\sum_{\lambda}\phi^\rho_\lambda(x)^* {\widetilde h}^{\bf k}_\lambda$
and 
\begin{eqnarray}
B_{\bf k}^-(0)&=&w_{\bf k}^* {g_{+-}(\ell,0:\omega_{kR})\over \Lambda_\pm(\omega_{kR})}
\ ,
\\
w_{\bf k}^*+\eta_\pm(\omega_{kR})B_{\bf k}^-(\ell)&=&
w_{\bf k}^* {1-\xi_{\pm}(\omega_{kR})g_{--}(0,0:\omega_{kR})\over \Lambda_\pm(\omega_{kR})}
\ .
\end{eqnarray}

Now we discuss the sign of small imaginary parts in the energy denominators. 
The sign should be chosen so that we have $e^{iH_{\rm MF}t/\hbar} a_{{\bf k}\sigma} e^{-iH_{\rm MF}t/\hbar} 
e^{i\omega_{kL} t}\to
\alpha_{{\bf k}\sigma}$
and $e^{iH_{\rm MF}t/\hbar}b_{{\bf k}\sigma}e^{-iH_{\rm MF}t/\hbar}e^{i\omega_{kR} t}\to
\beta_{{\bf k}\sigma}$, ($t\to -\infty$). From (\ref{InAlpha}) and (\ref{InBeta}), original operators can be
expressed in terms of the incoming fields and e.g., 
\begin{eqnarray*}
&&e^{iH_{\rm MF}t/\hbar}a_{k\sigma}e^{-iH_{\rm MF}t/\hbar} e^{i\omega_{kL} t}-\alpha_{k\sigma} \nonumber\\
&&~~~~=v_{\bf k}^*
\integ{{\bf k}'}\Big\{
{{A_{{\bf k}'}^-(0)}^*\alpha_{{\bf k}'\sigma}e^{-i(\omega_{k'L}-\omega_{kL})t}
\over \omega_{k'L}-\omega_{kL}\pm i0}+
{{B_{{\bf k}'}^-(0)}^*\beta_{{\bf k}'\sigma}e^{-i(\omega_{k'R}-\omega_{kL})t}
\over \omega_{k'R}-\omega_{kL}\pm i0}
\Big\}
\end{eqnarray*}
which vanishes as $t\to -\infty$ only if the lower sign is chosen since 
$\displaystyle\lim_{t\to -\infty} {e^{-ixt}\over x+i0}=0$.

Then, the electron field in the TLM chain is given by
\begin{eqnarray}
\Psi_\sigma(x)&=&\sum_\lambda \phi_\lambda(x) f_{\lambda\sigma}=\sum_\lambda \phi_\lambda(x)
\int d{\bf k} 
\{h_\lambda^{{\bf k}*}\alpha_{{\bf k}\sigma}+{\widetilde h}_\lambda^{{\bf k}*}\beta_{{\bf k}\sigma}
\}\nonumber
\\
&=&\int d{\bf k} 
\Big\{
{v_{\bf k}\ \alpha_{{\bf k}\sigma} \over \Lambda_-(\omega_{kL})^*} h(x;\omega_{kL})+
{w_{\bf k}\ \beta_{{\bf k}\sigma} \over \Lambda_-(\omega_{kR})^*} {\widetilde h}(x;\omega_{kR})
\Big\}
\ ,
\end{eqnarray}
where
\begin{eqnarray}
h(x;\omega)&=& G(x,0;\omega)\binom{0}{1} \{1-g_{++}(\ell,\ell;\omega)\eta_+(\omega)\}
\nonumber\\
&&+G(x,\ell;\omega)\binom{1}{0} g_{+-}(\ell,0;\omega)\eta_+(\omega)
\ ,
\\
{\widetilde h}(x;\omega)&=&
G(x,0;\omega)\binom{0}{1} g_{-+}(0,\ell;\omega)\xi_+(\omega)
\nonumber\\
&&+G(x,\ell;\omega)\binom{1}{0} \{1-g_{--}(0,0;\omega)\xi_+(\omega)\}
\ .
\end{eqnarray}

\section{Coulomb energy and chemical potentials \label{ap.chem}}

Applying a bias voltage $V$ to the TLM chain corresponds to the 
prescription $\mu_L-\mu_R=-eV$ and it alone does 
not determine individual values of $\mu_L$ and $\mu_R$. 
However, as will be explained below, if $\mu_L+\mu_R\not=0$, the 
average number of electrons on the TLM chain increases or decreases 
and the whole system including ions would be electrically charged as 
compared with the equilibrium case. Then, such states have large 
electrostatic energy and are hard to be realized. Thus, one should
choose $\mu_L=-\mu_R=-eV/2$.

The proof is as follows. By a similar argument to the calculation
of $S$ in (\ref{SelfUni2}), the electron number density is found to be
\begin{eqnarray}
&&\sum_\sigma\left<\Psi_\sigma^\dag(x)\Psi_\sigma(x)\right>_\infty
\nonumber\\
&& ~=
\int_{\overline\Delta}^{\hbar \omega_c}{d\epsilon\over \pi\hbar v}{\epsilon
\over\sqrt{\epsilon^2-{\overline\Delta}^2}}
\ \big(f_L(\epsilon)+f_L(-\epsilon)+
f_R(\epsilon)+f_R(-\epsilon)
\big)
\ ,
\end{eqnarray}
provided that $x$ is not close to the chain ends.
At equilibrium where $\mu_L=\mu_R=0$, the sum of four Fermi distribution functions
is equal to two and, thus, irrespective to the temperature,
\begin{eqnarray}
\sum_\sigma\left<\Psi_\sigma^\dag(x)\Psi_\sigma(x)\right>_{\rm eq}
&=&
\int_{\overline\Delta}^{\hbar \omega_c}{d\epsilon\over \pi\hbar v}{2\epsilon
\over\sqrt{\epsilon^2-{\overline\Delta}^2}}
\ .
\end{eqnarray}
Then, as easily seen, one has
\begin{eqnarray}
&&\sum_\sigma\left<\Psi_\sigma^\dag(x)\Psi_\sigma(x)\right>_\infty
-\sum_\sigma\left<\Psi_\sigma^\dag(x)\Psi_\sigma(x)\right>_{\rm eq}
\nonumber\\
&&~=(1-e^{-(\mu_L+\mu_R)/T})
\int_{\overline\Delta}^{\hbar \omega_c}{d\epsilon\over \pi\hbar v}{\epsilon
\big(f_L(\epsilon)f_R(-\epsilon)+f_L(-\epsilon)f_R(\epsilon)
\big)
\over\sqrt{\epsilon^2-{\overline\Delta}^2}}
\ ,
\end{eqnarray}
which is nonzero unless $\mu_L=-\mu_R$.

In the rest of this appendix, we show that a number of electrons per site for the open SSH chain discussed
in Appendix~A is approximately unity by choosing  $\hbar\omega_c=\pi t_0$.
Indeed, since
\begin{eqnarray*}
\sum_\sigma \left<d_\sigma^+(x)d_\sigma(x)\right>
=\sum_\sigma \left<e_\sigma^+(x)e_\sigma(x)\right>
\frac{1}{\pi v\hbar}\int^{\hbar\omega_c}_{\hbar\bar{\Delta}_0}
d\epsilon\frac{\epsilon}{\sqrt{\epsilon^2-\bar{\Delta}_0^2}}
\approx\frac{\omega_c}{\pi v}
\end{eqnarray*}
and $v=2at_0/\hbar$, a number of electrons per site is
\begin{eqnarray*}
\sum_\sigma\left<C^+_{2n,\sigma}C_{2n,\sigma}\right>
=\sum_\sigma\left<C^+_{2n-1,\sigma}C_{2n-1,\sigma}\right>
\approx
\frac{\hbar\omega_c}{\pi t_0}=1 
\ .
\end{eqnarray*}

\section{Green function for spatially uniform phase \label{ap.Green}}

In this appendix, we explicitly write down the Green function defined by
(\ref{GreenFt1}) in the spatially uniform case:
\begin{eqnarray}
&&g_{++}(x,y;\omega)= 
\begin{cases} \displaystyle
-{\omega 
\big(\hbar\kappa v\cos \kappa(\ell-y)+{\overline\Delta}\sin \kappa(\ell-y)\big)\sin \kappa x
\over \kappa v^2D(\omega)} &(x<y)\cr\cr
\displaystyle
-{\omega 
\big(\hbar\kappa v\cos \kappa(\ell-x)+{\overline\Delta}\sin \kappa(\ell-x)\big)\sin \kappa y
\over \kappa v^2D(\omega)} &(x>y)\cr
\end{cases}
\\ 
\nonumber \\
&&g_{--}(x,y;\omega)= 
\begin{cases} \displaystyle
-{\omega 
\big(\hbar\kappa v\cos \kappa x+{\overline\Delta}\sin \kappa x\big)\sin \kappa (\ell-y)
\over \kappa v^2D(\omega)} &(x<y)\cr\cr
\displaystyle
-{\omega 
\big(\hbar\kappa v\cos \kappa y+{\overline\Delta}\sin \kappa y\big)\sin \kappa (\ell-x)
\over \kappa v^2D(\omega)} &(x>y)\cr
\end{cases}
\\ 
\nonumber\\
&&g_{+-}(x,y;\omega)=g_{-+}(y,x;\omega)\nonumber\\
&&~=\begin{cases} \displaystyle
-{\hbar\omega^2 
\ \sin \kappa (\ell-y)\ \sin \kappa x 
\over \kappa v^2D(\omega)} 
&(x<y)\cr\cr
\displaystyle
-{\big(\hbar\kappa v\cos \kappa y+{\overline\Delta}\sin \kappa y\big)
\big(\hbar\kappa v\cos \kappa (\ell-x)+{\overline\Delta}\sin \kappa (\ell-x)\big)
\over \hbar\kappa v^2D(\omega)} 
&(x>y)\cr
\end{cases}
\end{eqnarray}
where $\kappa=\sqrt{(\hbar\omega)^2-{\overline\Delta}^2}/(\hbar v)$ and 
$D(\omega)=\hbar\kappa v\cos \kappa \ell+{\overline\Delta}\sin \kappa \ell$.

\section{Stability of fixed points\label{ap.sta}}

In this appendix, we show that nontrivial solutions of (\ref{SelfUni1b}) are more stable 
at constant current than those at constant bias voltage. 
Then, we prove that, at constant current, the zero-temperature ordered phase 
with $\overline\Delta$ given by (\ref{ZeroPhase2}) is stable.

Firstly, we note that the stability indices $\chi_V$ at constant bias voltage and $\chi_I$ 
at constant current differ by
\begin{eqnarray}
\chi_I({\overline\Delta})-\chi_V({\overline\Delta})
&=&-\lambda{\overline\Delta}
\bigg(
{\partial S\over \partial V}
\bigg)_{\overline\Delta}
\bigg(
{\partial \overline{J}\over \partial {\overline\Delta}}
\bigg)_V \bigg/
\bigg(
{\partial \overline{J}\over \partial V}
\bigg)_{\overline\Delta}
\ .
\end{eqnarray}
As easily seen, we have
\begin{eqnarray}
{eT\over \sinh\Big({eV\over 2T}\Big)}
\pd{S}{V}{{\overline\Delta}}&=&\int_{\overline\Delta}^{\hbar\omega_c}
\frac{d\epsilon}{\sqrt{\epsilon^2-{\overline\Delta}^2}}
{e^2\sinh\Big({\epsilon\over T}\Big) \over \Big\{
\cosh\Big({eV\over 2T}\Big)+\cosh\Big({\epsilon\over T}\Big)
\Big\}^2}
>0
\nonumber
\\
\pd{\overline{J}}{V}{{\overline\Delta}}&=&
{G_0\over T}
\int_{\overline\Delta}^{\hbar\omega_c}  d\epsilon
\frac{\sqrt{\epsilon^2-{\overline\Delta}^2}}{\epsilon}
{1+\cosh\Big({eV\over 2T}\Big)
\cosh\Big({\epsilon\over T}\Big) \over \Big\{
\cosh\Big({eV\over 2T}\Big)+\cosh\Big({\epsilon\over T}\Big)
\Big\}^2}
>0
\nonumber
\\
{e {\overline\Delta}\over \sinh\Big({eV\over 2T}\Big)} 
\pd{\overline{J}}{\overline\Delta}{V}&=&
-
\int_{\overline\Delta}^{\hbar\omega_c}  d\epsilon
\frac{2G_0{\overline\Delta}^2}{\epsilon\sqrt{\epsilon^2-{\overline\Delta}^2}
\Big\{\cosh\Big({eV\over 2T}\Big)+\cosh\Big({\epsilon\over T}\Big)\Big\}}
<0
\ .
\nonumber
\end{eqnarray}
Thus, $\chi_I({\overline\Delta})>\chi_V({\overline\Delta})$ which implies that the phase is
more stable at constant current than at constant bias voltage.

Now, let us study the stability of the ordered phase given by (\ref{ZeroPhase2}) at constant current.
It is easy to show
\begin{equation}
{\overline\Delta}
\Bigg(
{\partial S\over \partial V}
\Bigg)_{\overline\Delta}
\Bigg(
{\partial \overline{J}\over \partial {\overline\Delta}}
\Bigg)_V \bigg/
\Bigg(
{\partial \overline{J}\over \partial V}
\Bigg)_{\overline\Delta}
={2r\over r^2-1}\Big(\int_1^rdx{\sqrt{x^2-1}\over x}-\sqrt{r^2-1}\Big)
\ ,
\end{equation}
where $r\equiv |eV/2{\overline\Delta}|>1$.
Thus, we obtain the desired result:
\begin{eqnarray}
\chi_I&=&
2\lambda\Big(
{\hbar\omega_c\over \sqrt{(\hbar\omega_c)^2-|{\overline\Delta}|^2}}-1\Big)
+{2\lambda r\over r^2-1}
\Big\{r-{1\over r}-
\int_1^rdx{\sqrt{x^2-1}\over x}\Big\}
\nonumber\\
&=&2\lambda\Big(
{\hbar\omega_c\over \sqrt{(\hbar\omega_c)^2-|{\overline\Delta}|^2}}-1\Big)
+{2\lambda r\over r^2-1}
\int_1^r dx
\Big\{1+{1\over x^2}-{\sqrt{x^2-1}\over x}\Big\}
\nonumber\\
&=&2\lambda\Big(
{\hbar\omega_c\over \sqrt{(\hbar\omega_c)^2-|{\overline\Delta}|^2}}-1\Big)
+{2\lambda r\over r^2-1}
\int_1^r dx
{1+3x^2 \over x^2(x^2+1+x\sqrt{x^2-1})}
>0 \ .
\nonumber
\end{eqnarray}

\section{Ginzburg-Landau expansion coefficients\label{ap.GL}}

In this appendix, we list up the coefficients $K_2$ and $K_4$ introduced in (\ref{GLExp}) when $\hbar\omega_c\gg T$.
By letting $\hbar\omega_c/T\to\infty$ in $S({\overline\Delta},V,T)-S(0,V,T)$ and Taylor-expanding the result 
with respect to ${\overline\Delta}/T$, we obtain the desired expansion:
\begin{eqnarray}
0={1\over 2\lambda}+{1\over 2}S(0,V,T)
+{S({\overline\Delta},V,T)-S(0,V,T)\over 2}
={\chi_N\over 2\lambda}-{K_2{\overline\Delta}^2\over 2T^2}
+{K_4{\overline\Delta}^4\over 8T^4}
\ ,
\end{eqnarray}
where $K_2$ and $K_4$ are functions of $eV/(2T)$ defined by
\begin{eqnarray}
K_2&=&\int_0^\infty {dt\over t}{d\over dt}\Big({\sinh t\over t(\cosh(eV/(2T))+\cosh t)}\Big)\ ,\nonumber\\
K_4&=&-
\int_0^\infty {dt\over t}{d\over dt}\Big\{{1\over t}{d\over dt} \Big({\sinh t\over t(\cosh(eV/(2T))+\cosh t)}\Big)\Big\}
\ .
\end{eqnarray}

\end{document}